\documentclass[useAMS,usenatbib]{mn2e}

\usepackage{graphicx}
\usepackage{amssymb}
\usepackage[authoryear]{natbib}

\bibpunct{(}{)}{;}{a}{}{,}

\title[Stellar population gradients in post-starburst galaxies]{Stellar population gradients and spatially resolved kinematics in luminous post-starburst galaxies} 
\author[Michael.~ B.~ Pracy et al.]{
\parbox[t]{\textwidth}{ Michael B.~Pracy$^{1}$,  Scott Croom$^{1}$, Elaine Sadler$^{1}$, Warrick J.~Couch$^{2}$, 
Harald Kuntschner$^{3}$, Kenji Bekki$^{4}$,  Matt S.~Owers$^{5}$, Martin Zwaan$^{3}$, James Turner$^{6}$, Marcel Bergmann$^6$ }\\
\vspace*{6pt}\\
$^1$Sydney Institute for Astronomy, School of Physics, University of Sydney, NSW 2006, Australia\\
$^2$Center for Astrophysics \& Supercomputing, Swinburne University of Technology, P.O. Box 218, Hawthorn, Vic, Australia\\
$^3$European Southern Observatory, Karl-Schwarzschild Strasse 2, 85748, Garching, Germany\\
$^4$ICRAR, M468, The University of Western Australia, 35 Stirling Highway, Crawley Western Australia 6009, Australia\\
$^5$Australian Astronomical Observatory, P.O. Box 296, Epping, NSW 1710, Australia \\
$^6$Gemini Observatory, Casilla 603, La Serena, Chile\\
}

\begin{document}

\date{Received 0000; Accepted 0000}

\pagerange{\pageref{firstpage}--\pageref{lastpage}} \pubyear{2010}

\maketitle

\label{firstpage}

\begin{abstract}
We have used deep integral field spectroscopy obtained with the GMOS instrument on Gemini-North to determine the spatial distribution 
of the post-starburst stellar population in four luminous ($\sim L^{*}$) E+A galaxies at $z<0.04$.  We find all four 
galaxies have centrally-concentrated gradients in the young stellar population contained within the central $\sim 1$\,kpc. 
This is in agreement with the Balmer line gradients found in local low luminosity E+A galaxies. The
results from higher redshift ($z\sim 0.1$) samples of luminous E+A galaxies have been varied, but in general 
have found the post-starburst signature to be extended or a galaxy-wide phenomenon or have otherwise failed to detect gradients in
the stellar populations. The ubiquity of the detection of a centrally concentrated young stellar population in local 
samples, and the presence of significant radial gradients in the stellar populations when the E+A galaxy core 
is well resolved raises the possibility that spatial 
resolution issues may be important in interpreting the higher redshift results. The two early type E+A galaxies in our sample
that can be robustly kinematically classified, using the $\lambda_{R}$ parameter, are fast--rotators. 
Combined with previous measurements, this brings the total number of E+A galaxies with measurements of $\lambda_{R}$
to twenty-six, with only four being classified as slow--rotators. This fraction is similar to the fraction of
the early--type population as a whole and argues against the need for major mergers in the production of E+A galaxies, since
major mergers should result in an increased fraction of slow rotators.
\end{abstract}

\begin{keywords}
galaxies: evolution -- galaxies: stellar content
\end{keywords}

\section{Introduction}
E+A galaxies are in the process of a dramatic change in their star formation properties. They are characterized by an
optical spectrum devoid of emission lines but exhibiting strong Balmer absorption
lines. The strong Balmer absorption is associated with the presence of a significant 
population of A-type stars which must have formed within the last $\sim 1$\,Gyr, yet the absence 
of emission lines implies that star formation is not ongoing. This spectral signature is interpreted
as that of a post-starburst galaxy which has undergone a burst of star formation that has been
truncated abruptly \citep[e.g.][]{couch87,poggianti99}. These galaxies are undergoing a colour evolution
between the blue cloud and the red sequence, and this evolution may be accompanied by a morphological transformation
where a star-forming disk galaxy is transformed into a quiescent spheroidal 
system \citep[e.g.][]{caldwell96,zabludoff96}.

There has been a variety of environmental mechanisms suggested which could cause these rapid changes in the star formation
rate of a galaxy. These include both major and minor galaxy mergers \citep{mihos96,bekki05}, gravitational interactions between
galaxies of close to equal mass \citep{bekki01}, and in denser environments: galaxy harassment \citep{moore96}, interaction with
the intra-cluster medium and tidal interactions with the global cluster field \citep{bekki01}. While the situation in galaxy clusters
remains unclear \citep{pracy10} and likely requires a cluster specific mechanism \citep{tran04,pracy05,poggianti09}, there is 
evidence that galaxy mergers are responsible for producing much of the local field E+A population. This evidence includes an increased rate
of tidal features \citep{zabludoff96,yang08,pracy09}, the nature of the clustering and luminosity function \citep{blake04},
and the internal kinematics from spatially resolved spectroscopy which favors a scenario dominated by minor mergers \citep{pracy09,pracy12}.
\citet{goto05} found an increased rate of near neighbors for E+A galaxies in the Sloan Digital Sky Survey (SDSS) implicating tidal interactions
as a formation mechanism.

If gas-rich mergers are responsible for the E+A population then evidence of this should be imprinted in the 
spatial distribution of the stellar populations. During a merger, tidal forces transfer 
angular momentum from the gas to the stars \citep{barnes96}, funneling the gas into the galactic centre and producing 
a central starburst \citep[e.g.][]{hopkins09}.  
Furthermore, feedback processes cut off the star formation \citep{springel05} leaving a young stellar population in a centrally concentrated cusp
and an old stellar population distributed like a normal early type galaxy \citep{hopkins09,snyder11}.
The young component should be compact, with scales of $\sim$1\,kpc \citep{bekki05,hopkins09}.
In the E+A galaxies this young component  should be observable as a Balmer line absorption enhancement and gradient 
in the central region \citep{pracy05,snyder11}.

However, from an observational standpoint there remain conflicting results regarding the spatial distribution of the post-starburst population 
and the presence of stellar population gradients. \citet{norton01} followed up a sample of E+A galaxies selected from the Las Campanas 
Redshift Survey \citep[LCRS,][]{zabludoff96}
with long-slit spectroscopy. They compared the radial distribution of the young and old stellar populations and concluded that the young stellar populations
were centrally concentrated with respect to the old population. The seeing FWHM of the \citet{norton01} observations corresponded to $\sim$2--3~kpc and in approximately 
half of the cases they were able to spatially resolve the young stellar population. This lead them to conclude that 
in these cases the young stellar population is not confined to the galaxy core. \citet{chilingarian09} obtained Integral Field Unit (IFU) spectroscopy of  
a gas rich post-starburst galaxy
and found the young population to be spread over the central $\sim 2.5$\,kpc. Likewise \citet{yagi06} observed three E+A galaxies 
selected from the SDSS with long slit spectroscopy
and concluded that the young stellar population was as extended as the continuum light. \citet{pracy09} obtained Gemini Multi-Object Spectrograph (GMOS) IFU 
spectroscopy of a sample of ten E+A galaxies selected from the 2dFGRS \citep{blake04} with a median redshift of $z\sim 0.1$, and were unable to 
detect any Balmer line gradients or central concentration in the young population. However, their limited spatial coverage 
coupled with spatial resolution constraints imposed by the delivered image quality meant they could not robustly rule out spatial gradients. \citet{swinbank12} who
used the GMOS IFU observed a sample of eleven E+As selected from the SDSS and found that the A-stars extended over 2-15 kpc$^2$ or around one third of the galaxy area 
defined by the continuum light. This was based on the area of the galaxy exhibiting an equivalent width of the 
H$\delta$ line $> 6$\AA\, (representing a very stringent constraint). They concluded that  the characteristic E+A signature 
is a property  of the galaxy as a whole and not due to a heterogeneous mixture of populations.

These attempts to measure the spatial distribution and radial gradients in the stellar populations have been hampered by physical scale resolution constraints.
Since E+A galaxies are rare in the local universe, samples for detailed follow up are selected from large redshift surveys  such
as the SDSS and 2dFGRS \citep{blake04,goto07}. These surveys have median redshifts of $z\sim 0.1$ and as a result the majority of 
E+A galaxies followed up with spatially resolved spectroscopy are at a similar redshift. At this redshift 1\,kpc projects to $\sim 0.5$\arcsec, meaning
that the angular size of a typical ground--based seeing profile is significant in comparison to both the expected size of the gradients ($\sim 1$\,kpc) and
the normal scale length of galaxies (a few kpc). The observed distribution of stellar populations is a complex
combination of the delivered image quality (seeing), the true spatially resolved distribution of the stellar populations, and importantly (but sometimes overlooked)
the spatial distribution of the continuum light (i.e. the galaxy surface brightness profile). Detailed modeling of the convolution of the expected gradients with 
the seeing disk for E+A galaxies by \citet{pracy10} demonstrated that this can lead to misleading and ambiguous results.

\citet{pracy12} recently circumvented this problem by slightly modifying the E+A selection criteria to select a 
lower redshift sample of galaxies than was available in the literature ($z<0.01$).  This low redshift selection meant that the galaxies 
have a large angular to physical scale which allowed the central $\sim$1\,kpc region 
of the galaxies to be resolved; the region where stellar population gradients are expected. 
IFU spectroscopy of this sample enabled the first robust detections of Balmer line gradients in the centres of E+A galaxies. 
Six out of the sample of seven, and all the galaxies with regular morphologies, were observed to have compact and centrally concentrated 
Balmer absorption line gradients. 

An unavoidable consequence of the local selection of the \citet{pracy12} sample is that the sample is 
composed entirely of intrinsically faint galaxies an order of magnitude or more below the characteristic 
galaxy luminosity (L$^{*}$) and fundamentally different to the populations studied at higher redshift.
The physical effects of mergers, tidal interactions and shocking on star formation are notably different for 
dwarf galaxies than massive galaxies \citep[e.g.][]{icke85,brosch04,smith10} and the \citet{pracy12} results should not
be generalized to classical massive E+A galaxies. In addition, as a result of aperture effects, all but one of this
sample would not have been classified as E+A galaxies if observed beyond z$\sim 0.06$ \citep{pracy12}.

What is required is a sample of $\sim L^{*}$ galaxies 
at a low enough redshift where the physical scale resolution is still adequate to resolve stellar population
gradients should they be present. There does exist a `sweet spot' at $z\lesssim 0.04$ which allows a large enough 
angular to physical scale to resolve these gradients given good ground--based seeing and enough cosmic volume 
to select a small sample of L$^{*}$ targets. In this paper we present a study of the spatially resolved stellar populations in a sample of
four such E+A galaxies, based on IFU spectroscopy obtained with the Gemini Multi-Object Spectrograph.

\section{Observations and data}

\subsection{Sample selection}
\begin{figure}
     \includegraphics[width=5.8cm, angle=90, trim=0 0 0 0]{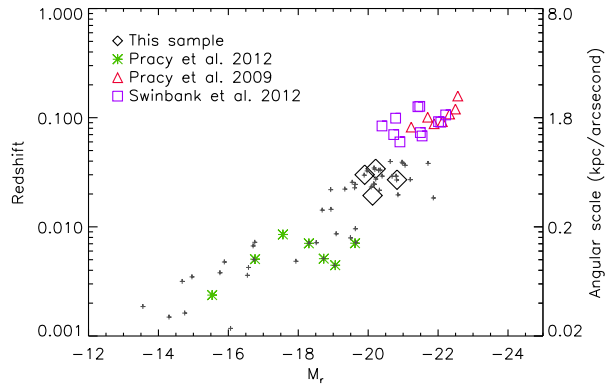}
\caption{\label{fig:magscale} Comparison of the earlier IFU E+A samples with the sample presented here. Absolute r-band magnitude plotted against redshift ({\it left axis}) and 
physical to angular scale in kpc per arcsecond ({\it right axis}).
This sample ({\it black diamonds}) consists of galaxies intrinsically brighter than the local sample of \citet[][{\it green stars}]{pracy12} and similar in 
brightness to the fainter galaxies in the higher redshift samples of \citet[][{\it red triangles}]{pracy09} and \citet[][{\it blue squares}]{swinbank12} but with significantly 
better physical scale resolution. The small {\it grey crosses} are the parent sample from which our four targets were selected based on the criteria in Section 2.1.}
\end{figure}

We obtained Gemini-North GMOS IFU spectroscopy of four E+A galaxies in the 2006B semester (Program ID: GN-2006B-Q-48; PI: Turner). This
small pilot sample was selected from a parent catalogue of E+A galaxies generated from the SDSS data release 4 database. The parent 
sample was restricted to objects with z$<$0.04 and red magnitudes $r<16.0$\,mags. The objects were 
selected from the SDSS to have strong H$\delta$ absorption and significant absorption when averaged over the 
H$\delta$, H$\gamma$, and H$\beta$ lines as well as little or no H$\alpha$ emission. Explicitly the selection constraints were:
\begin{list}{$\bullet$}{\itemsep=0.1cm}

\item EW(H$\alpha$)$> -2.5$\,\AA

\item EW(H$\delta$)$> 4.0$\,\AA

\item average of EW(H$\delta$), EW(H$\gamma$), EW(H$\beta$) $> 4.0$\,\AA

\item $z<0.04$

\item $r<$16\,mag

\end{list}
The sample defined by these criteria is shown as the {\it grey crosses} in Figure \ref{fig:magscale} -- our target galaxies have absolute magnitudes around the bright-end of
this sample. The redshifts and magnitudes of our targets translate to all being within 1.4 magnitudes of $M_{r}^{*}$. This coincides with the low end of the
luminosity distribution of the higher redshift IFU E+A samples of \citet{pracy09} and \citet{swinbank12}, but with a factor of $\sim$4 better physical to angular scale.
They are significantly more luminous than the most luminous galaxies in the local sample of \citet{pracy12}; see Figure \ref{fig:magscale} for a comparison. 
The sample consists of two isolated
early type galaxies (classified by visual inspection of the SDSS imaging), one early type cluster 
galaxy at a projected cluster-centric distance of $\sim 1.3$\,Mpc, and one strongly interacting system.

\subsection{Observations}
The data are composed of IFU observations taken in 2-slit mode giving a $5\times 7$ arcsecond
field-of-view with $\sim 0.2$\arcsec\, sampling and were taken in sub-arcsecond seeing (see Table \ref{tab:targets}). 
The B600 grating was used resulting in spectra with a resolution of $\sim 1688$ 
and covering a wavelength range from $\sim 4000$ to 5200\AA\,, which includes the H$\delta$, H$\gamma$ and H$\beta$ spectral lines.  Each science target was observed either
three or four times over two nights with each exposure being approximately 1\, hour in length. The exposures
are dithered both spatially and spectrally and for each exposure a corresponding flat field and arc lamp exposure is acquired
immediately preceding or following each science exposure. In addition, twilight flat fields in each grating position were taken
on the first night on which science observations for this project were made. A summary of the targets and observations is given in Table \ref{tab:targets}.
\begin{table*}
\begin{center}
\caption{\label{tab:targets}Summary of sample and observations}
\begin{tabular}{ccccccccccl}\hline
object ID          & $r$    &  $g-r$     &  $z$   &  $\rm{M}^{*} - \rm{M}_r$  & scale        &  H$\delta\gamma\beta$ & H$\alpha$      & exposure & seeing    & Morphology and      \\ 
                        & (mag)  &  (mag)     &        &    (mag)            & (kpc/arcsec) &        (\AA)               &  (\AA)     &    (s)    & (arcsec) &     environment notes              \\ \hline
SJ1613+5103      &  15.64 &  0.27      & 0.034  &      -0.98               &     0.68       &         6.36               &  -0.49     &   14160    &   0.83   & Strongly interacting pair \\ 
SJ2114+0032      &  14.54 &  0.33      & 0.027  &      -0.39               &     0.54       &         7.09               &  0.87      &   14160    &   0.61   & Isolated Field                     \\
SJ1718+3007      &  15.70 &  0.62      & 0.030  &      -1.32               &     0.60       &         5.51               & -0.39      &   10620    &   0.81   & Isolated Field  \\
SJ0044-0853      &  14.50 &  0.58      & 0.019  &      -1.09               &     0.39       &         4.17               &  2.49      &   14160    &   0.66   &  $\sim$1.3\,kpc from Abell 085 centre \\ \hline 
\end{tabular}
\begin{flushleft}
Summary of the sample properties and observations. From {\it left to right:} object ID; r-band magnitude; $g-r$ color; redshift; the difference between
the absolute r-band magnitude and the characteristic magnitude of the r-band luminosity function from \citet{montero09}; physical to angular scale; average equivalent width
of the H$\delta$, H$\gamma$, and H$\beta$ lines from the SDSS spectra; equivalent width of the H$\alpha$ line from the SDSS spectra; total exposure time;
the mean seeing estimated from the acquisition images; environmental classification of the target galaxy. Note: object IDs are shortened for convenience, full SDSS 
object IDs are ({\it top to bottom}): 587729227690475600, 587730847963545655, 587729408621609096, 587727227842461731.
\end{flushleft}
\end{center}
\end{table*}

\subsection{Data reduction}
The data were reduced using the standard {\sc iraf} routines from the {\sc gemini} package
which are specifically designed for reduction of GMOS data. Briefly, the fibre flat fields were traced and extracted and these
fibre positions were used for the extraction of the corresponding science and calibration frames. The twilight flat field was
extracted and used together with the flat field to make a response curve for each fiber using the task {\sc gfresponse}. The
arc frame was then extracted and the task {\sc gswavelength} used to establish a wavelength solution. The science frames were then bias--subtracted,
extracted, flat fielded and wavelength calibrated. Sky subtraction was performed for the two slits independently. Data cubes were produced using
the task {\sc gfcube} with 0.1\arcsec spatial sampling. The shifts between exposures as a result of the dither pattern were measured and
the separate data cubes for each target galaxy coadded using the task {\sc scombine}. Following this the data cubes were rebinned to 0.2\arcsec
spatial pixels. No flux calibration of the data was applied. We flattened out any
continuum slopes by division of a fitted low order polynomial. This is of little consequence since the analysis is restricted to relative line 
index measurements which, to first order, have no dependence on the overall continuum slope. In addition, the wavelength range of the observations 
is narrow (of the order $\sim 1000$\AA) and the galaxy continuum is expected to be relatively flat over that range.

\begin{figure*}
      \includegraphics[width=17.8cm, angle=0, trim=0 0 0 0]{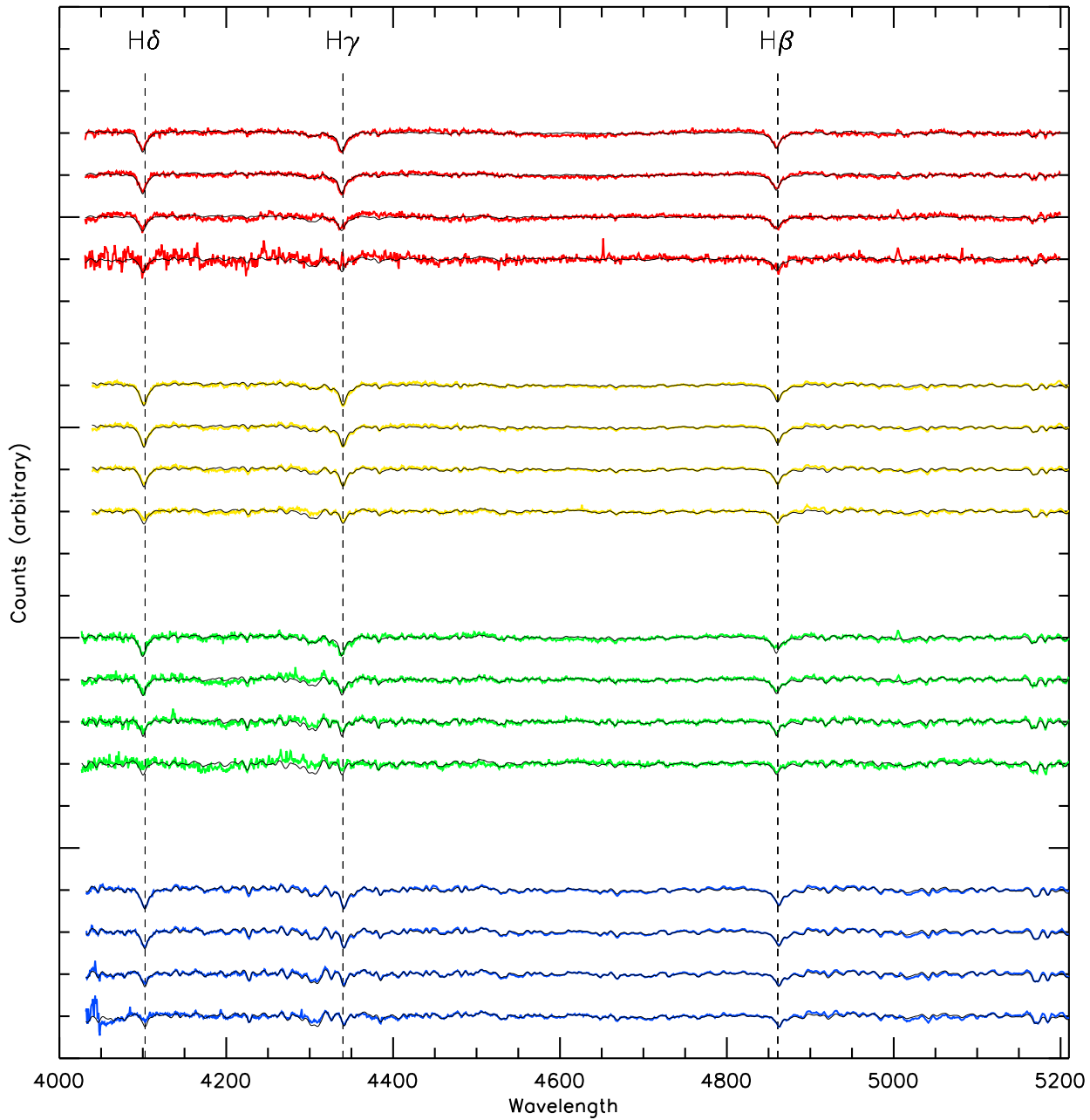}
\caption{\label{fig:spec}The annular binned spectra of each target galaxy plotted in different colours: SJ1613+5103 ({\it red}); SJ2114+0032
({\it yellow}); SJ1718+3007 ({\it green}); SJ0044-0853 ({\it blue}). For each galaxy the spectra are ordered from the innermost ({\it top}) to 
outermost ({\it bottom}) annular bin. The black over-plotted lines are the stellar population fits described in Section \ref{sec:fit}. The continuum 
shape has been flattened by division of a low order polynomial fit.}
\end{figure*}

As a result of the steep surface brightness profile of our target galaxies, the signal-to-noise ratio varies significantly with position on 
the IFU, generally in the sense that the signal-to-noise decreases rapidly with galacto-centric radius. 
In order to detect radial gradients in the stellar populations the spatial resolution and sampling in the innermost regions is critical. 
We do no binning or smoothing within a central radius of
1.5\arcsec (note the seeing full width half maximum of the data in all cases is $\sim$0.6--0.8 arcseconds). At larger galacto-centric radius, improvements
to the signal-to-noise ratio are required and we boxcar smooth the data to 0.6\arcsec (i.e. $3\times 3$ spaxels) to improve the signal-to-noise ratio in
these regions. In addition to the rapidly decreasing surface brightness profile of the target galaxies, the spatial edges of the IFU data have not had the total exposure 
time as a result of the dither pattern and we clip the edges of the IFU field to a final size of 5.2\arcsec$\times $4\arcsec.
Since we are primarily interested in radial trends in the stellar populations, we also construct annular binned spectra with inner and outer 
boundaries at radii of 0.0--0.4, 0.4--0.8, 0.8--1.6, 1.6--2.4 arcseconds. Note: since we use the same angular size for galaxies at different redshifts, the annular bins 
correspond to different physical apertures. 
This binning produces high quality spectra for radial analysis. The annular spectra for each galaxy are shown in Figure \ref{fig:spec}

\subsection{\label{sec:fit} Recession velocities and spectral fitting}
To study the internal kinematics and for a wavelength reference for measuring line indices,
we need to fit the recession velocity and velocity dispersion for each spaxel. We fitted the spectra
with a set of rest-frame single-age single-metallicity stellar population templates \citep{vazdekis10}.
The galaxy spectra were smoothed to match the resolution of the templates. The fitting was performed
using the penalized pixel fitting algorithm of \citet{Cappellari04}. A low order multiplicative polynomial is  
included in the fit to allow for variations in the continuum slope.

\section{Results}

\subsection{Balmer line gradients}
Our primary motivation is to measure the Balmer line strength distribution at high spatial resolution.
To this end we measured the equivalent width of the H$\delta$, H$\gamma$, and H$\beta$ lines for every
spaxel in our IFU data cube. The equivalent widths were measured using the standard flux summing technique. The line strengths 
were measured using the  Lick/Intermediate Dispersion Spectrograph system line definitions \citep{worthey97,trager98}, which included
smoothing of the spectra to Lick resolution. We use the broader H$\delta_{\rm A}$ and H$\gamma_{\rm A}$ index definitions throughout. 
The numerical values of the measured equivalent widths have a small  dependence 
on the continuum shape. In order to make our equivalent width measurements numerically consistent with the Lick system for comparison to models
we applied the following corrections: For each science spectrum we used the results of the stellar 
population fits described in Section \ref{sec:fit} (which included a multiplicative polynomial to remove
the dependence of the fit on continuum shape) to reconstruct  a `flux calibrated' model spectrum from the best fitting combination of
stellar population templates. This spectrum should have approximately the true continuum shape. The overall shape of this spectrum 
and the science spectrum were fitted with a low order polynomial and the science spectrum was multiplied by the ratio of these to give it the expected
`flux calibrated' shape. We then measured the index values from the `pseudo-flux-calibrated' spectrum.
The correction between measurements made on flux calibrated spectra and the Lick system are known \citep{norris06} and we applied this correction to
the measured equivalent width values to obtain final index values numerically consistent with the Lick system. While these corrections are required
for consistency, the dependence of the measured equivalent widths on the continuum shape is a second order effect, and 
the resulting corrections are small and do not qualitatively change the results.

\begin{figure*}
  \begin{center}
    \begin{minipage}{0.95\textwidth}
      \includegraphics[width=2.9cm, angle=0, trim=0 0 0 0]{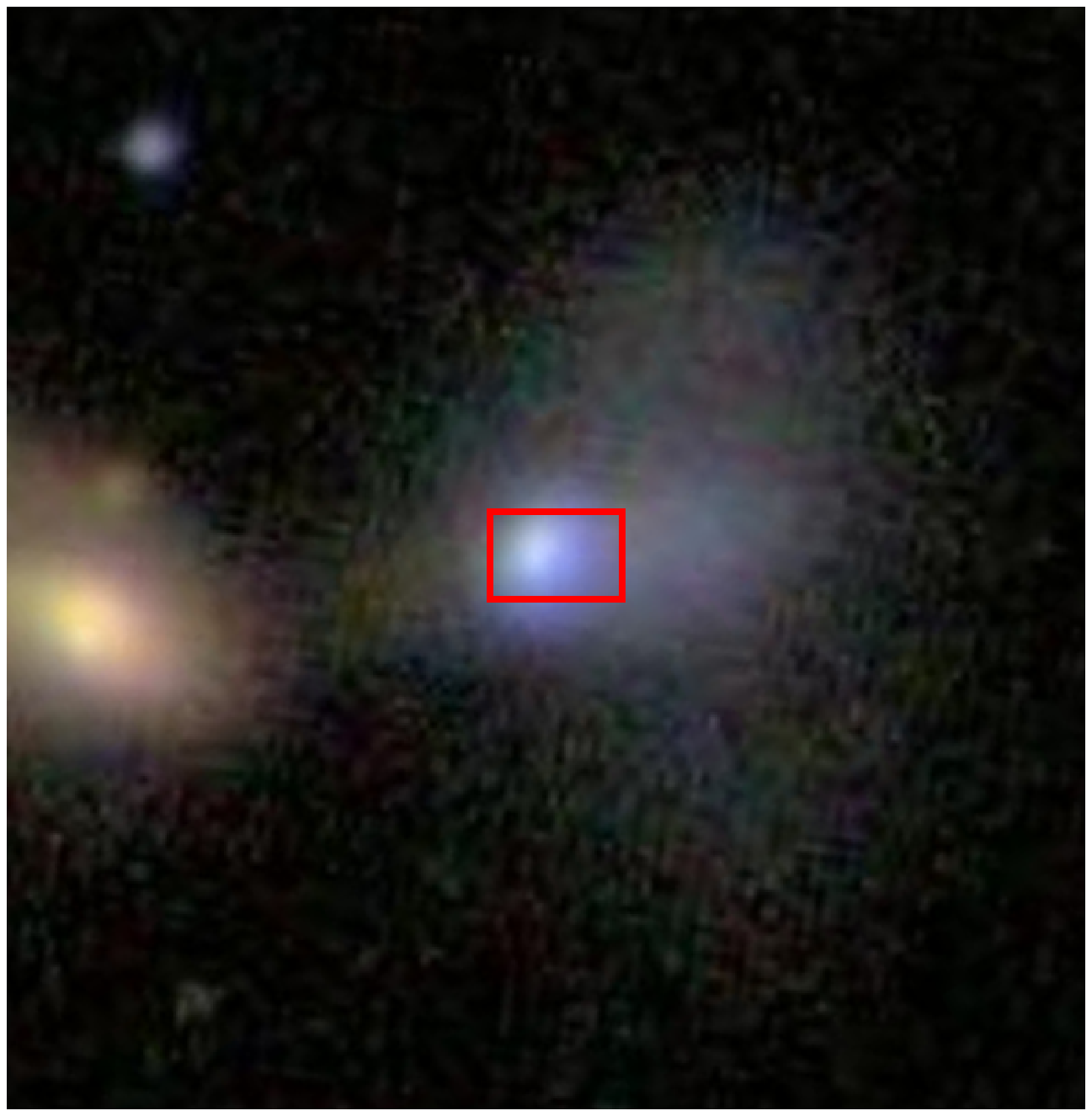}
      \includegraphics[width=5.1cm, angle=0, trim=20 35 0 0]{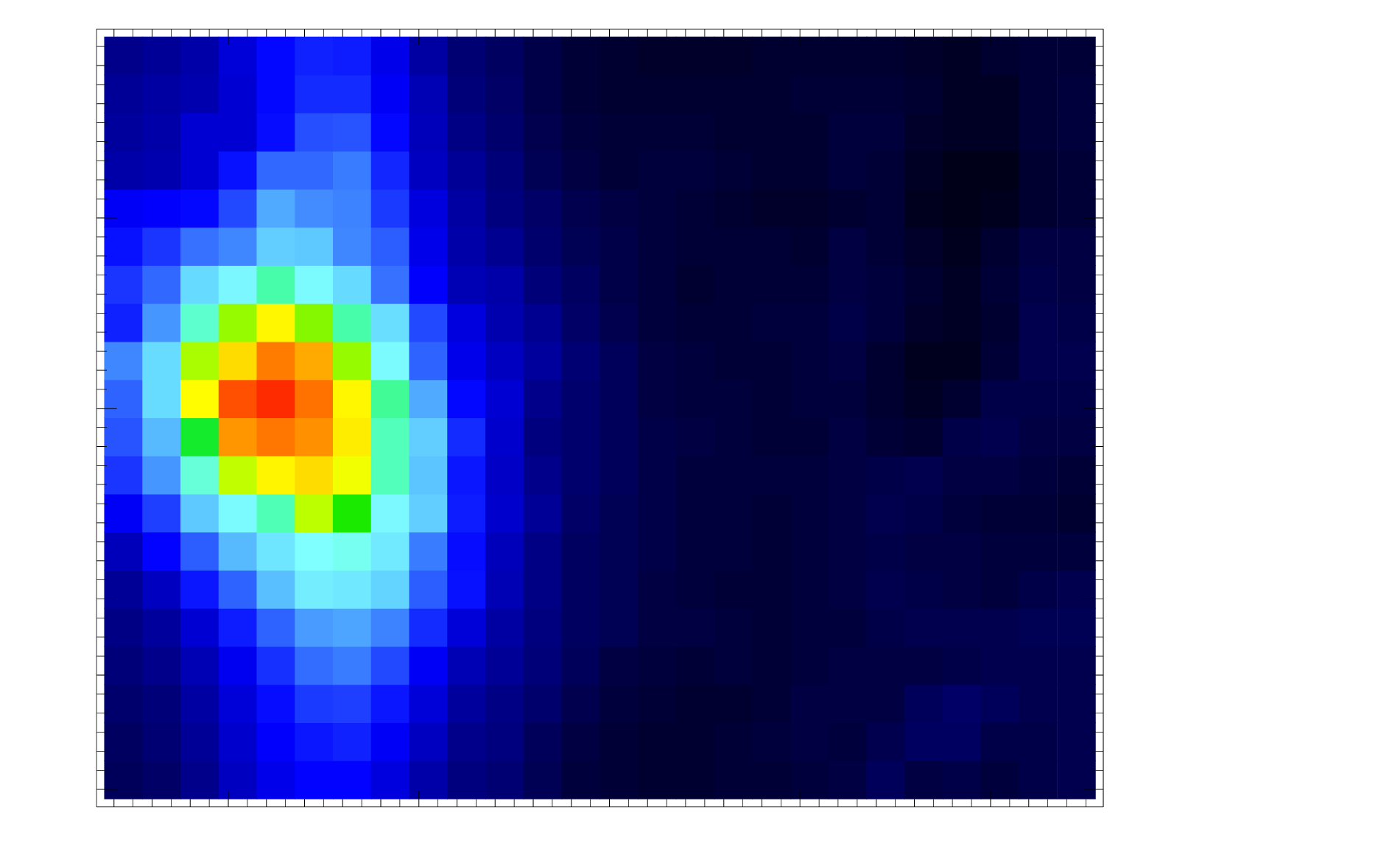}
      \includegraphics[width=4.0cm, angle=0, trim=200 35 0 0]{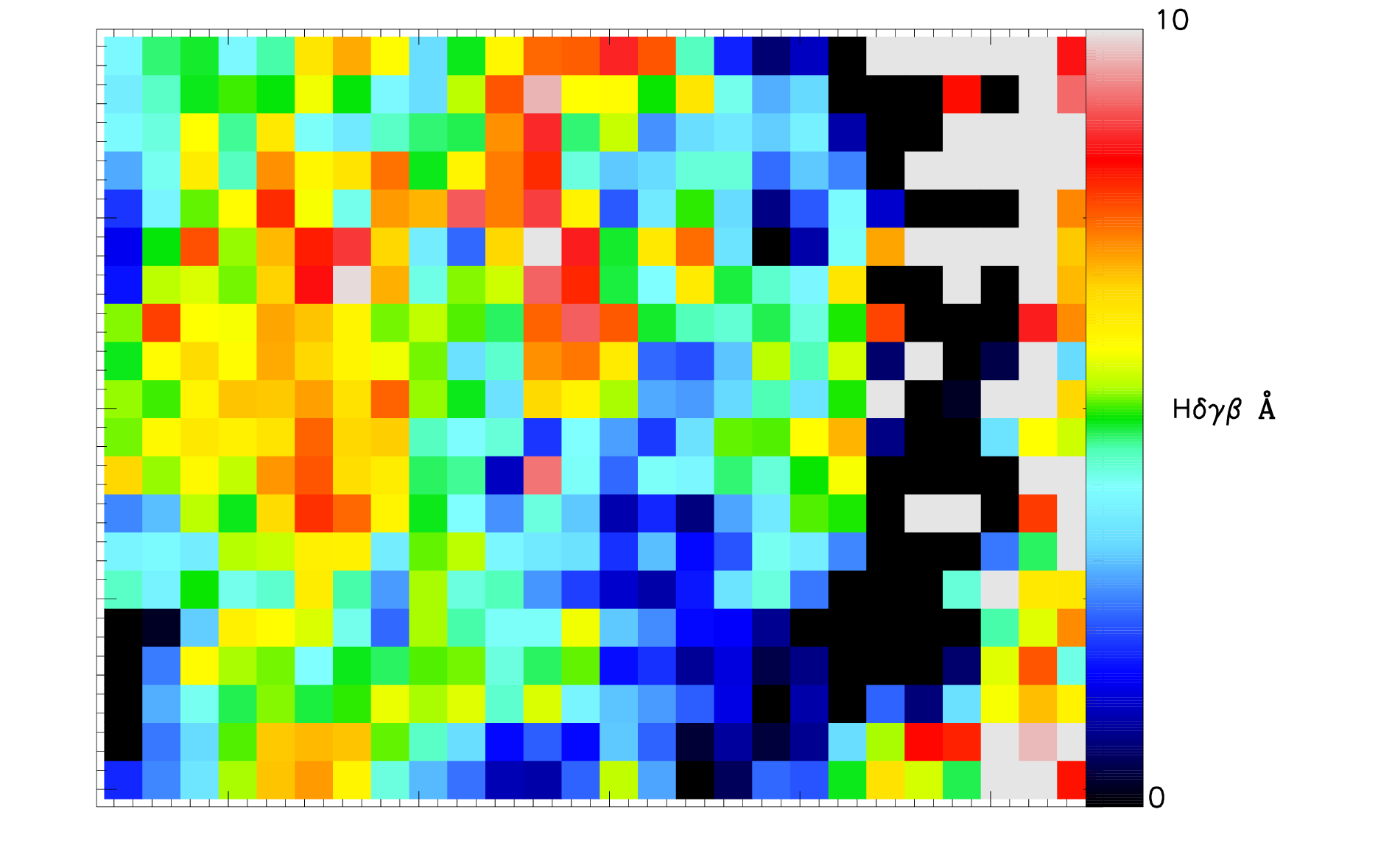}
      \includegraphics[width=3.1cm, angle=90, trim=60 0 0 100]{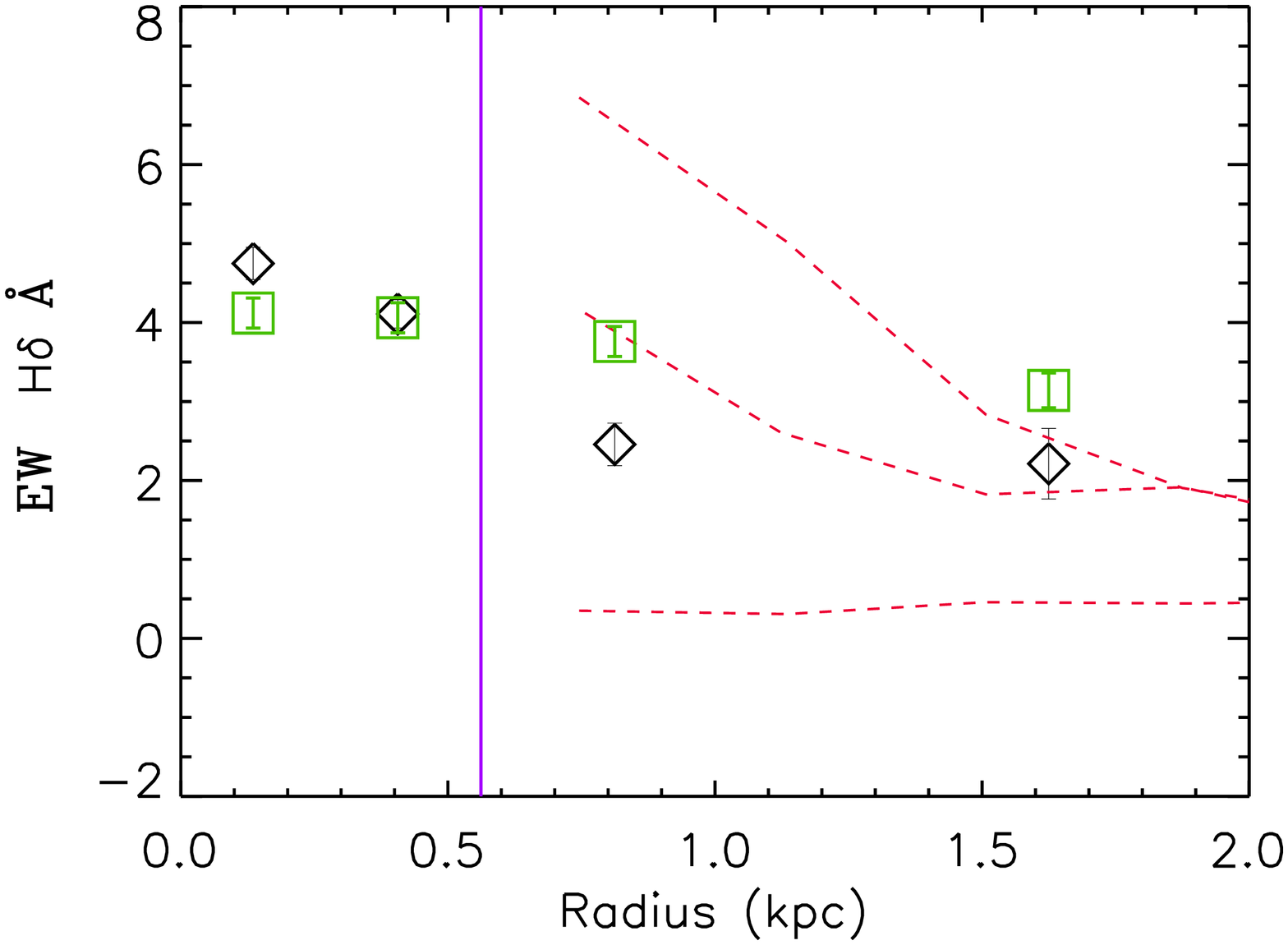}      
    \end{minipage}
    \begin{minipage}{0.95\textwidth}
      \includegraphics[width=2.9cm, angle=0, trim=0 0 0 0]{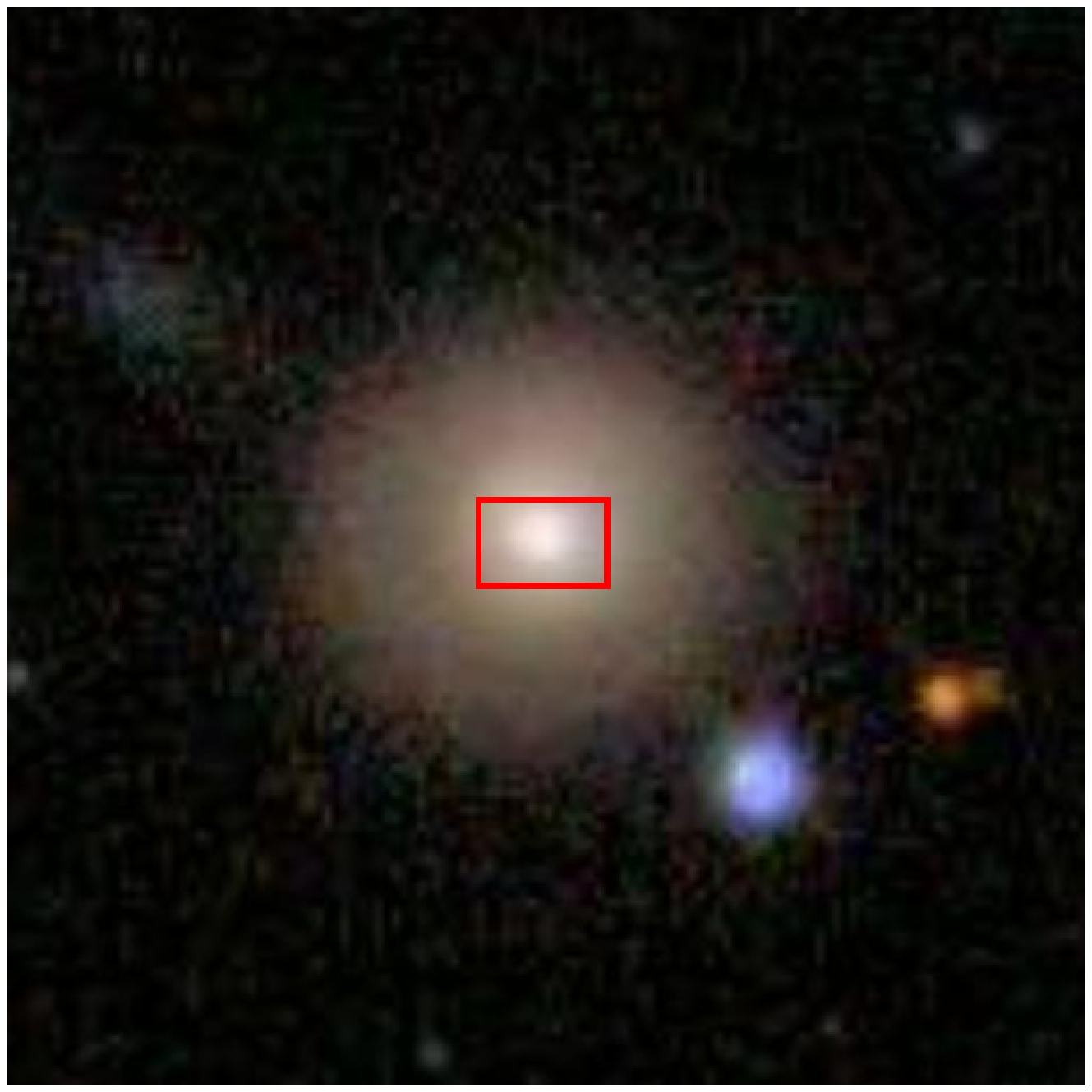}
      \includegraphics[width=5.1cm, angle=0, trim=20 35 0 0]{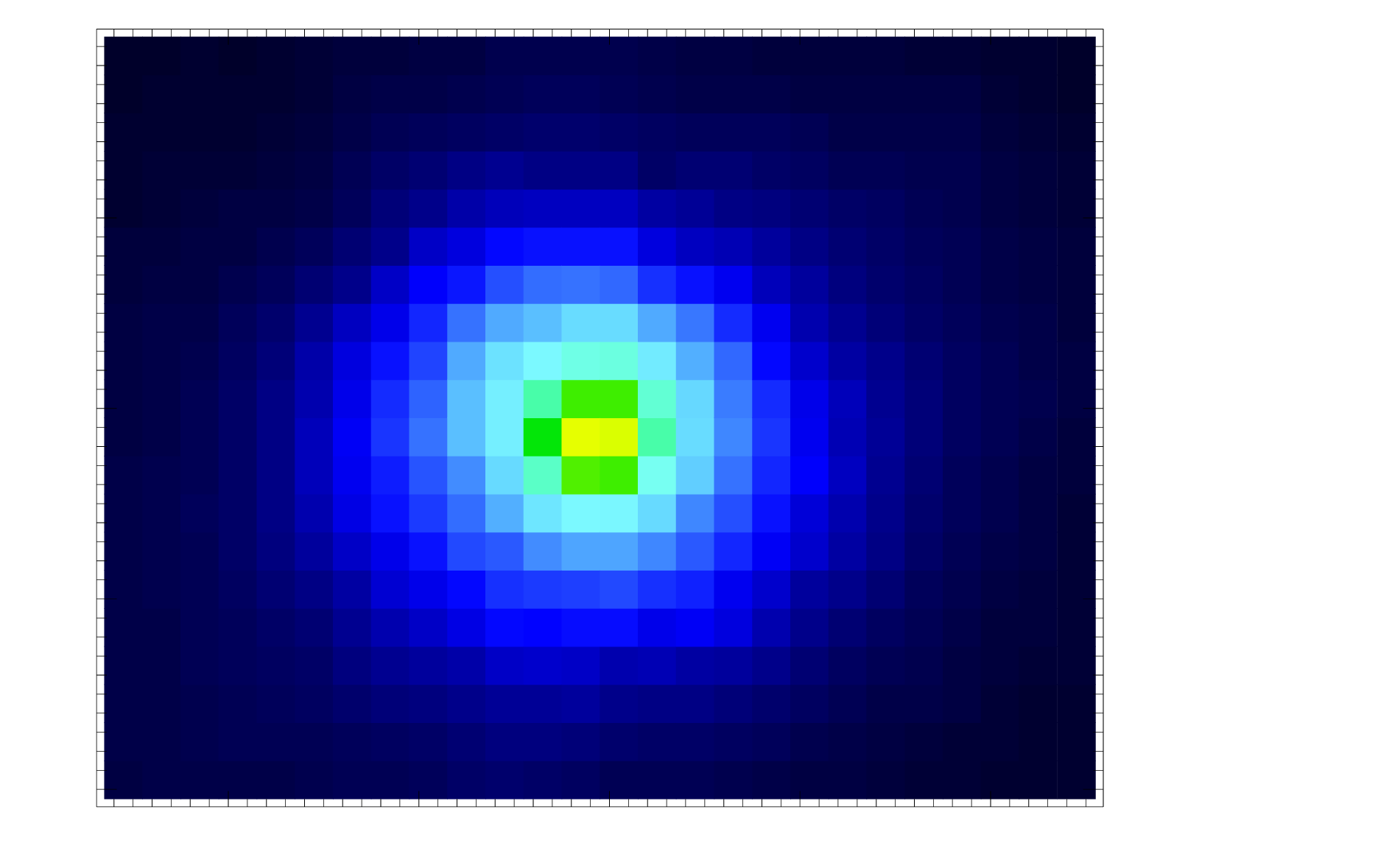}
      \includegraphics[width=4.0cm, angle=0, trim=200 35 0 0]{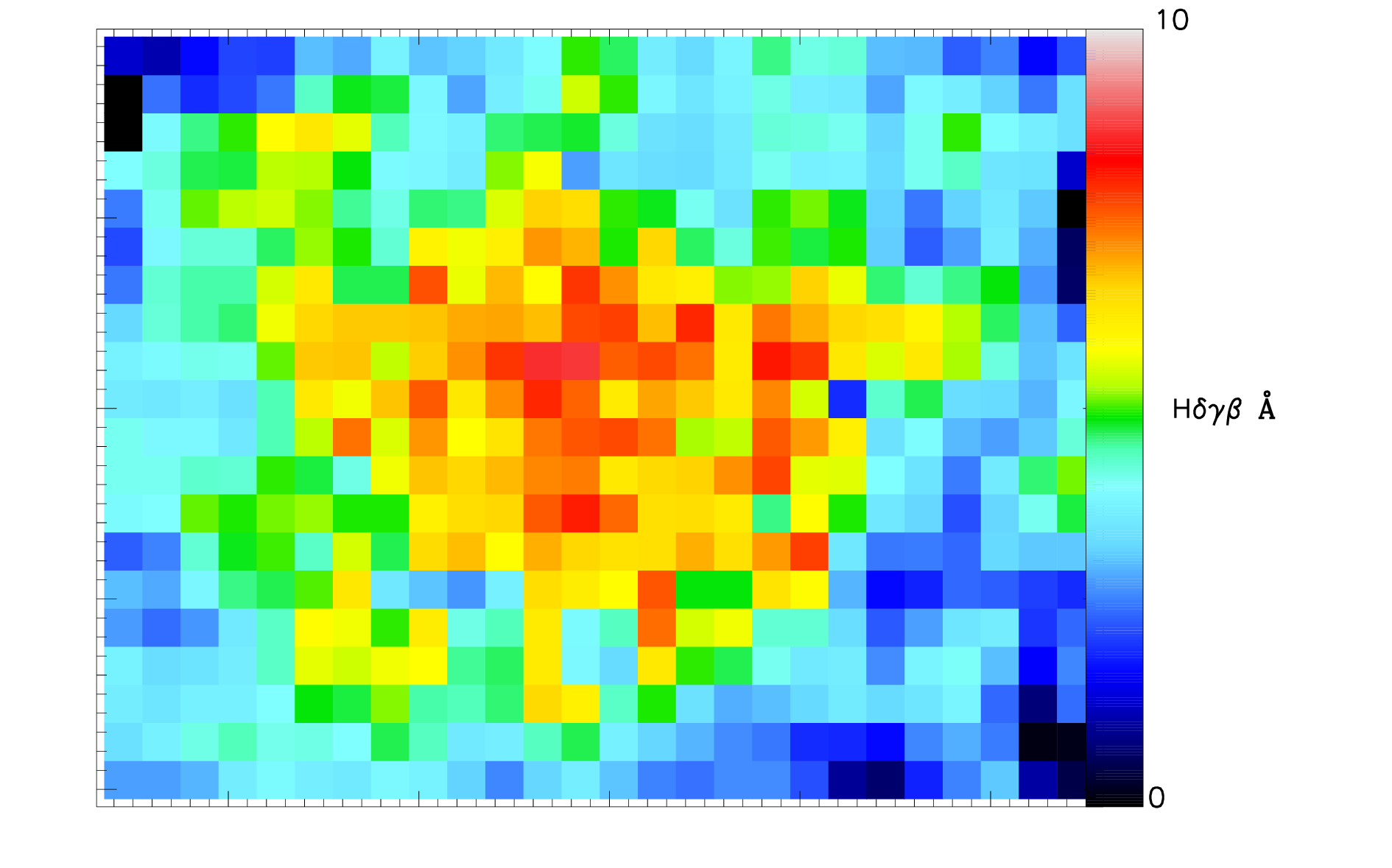}
     \includegraphics[width=3.1cm, angle=90, trim=60 0 0 100]{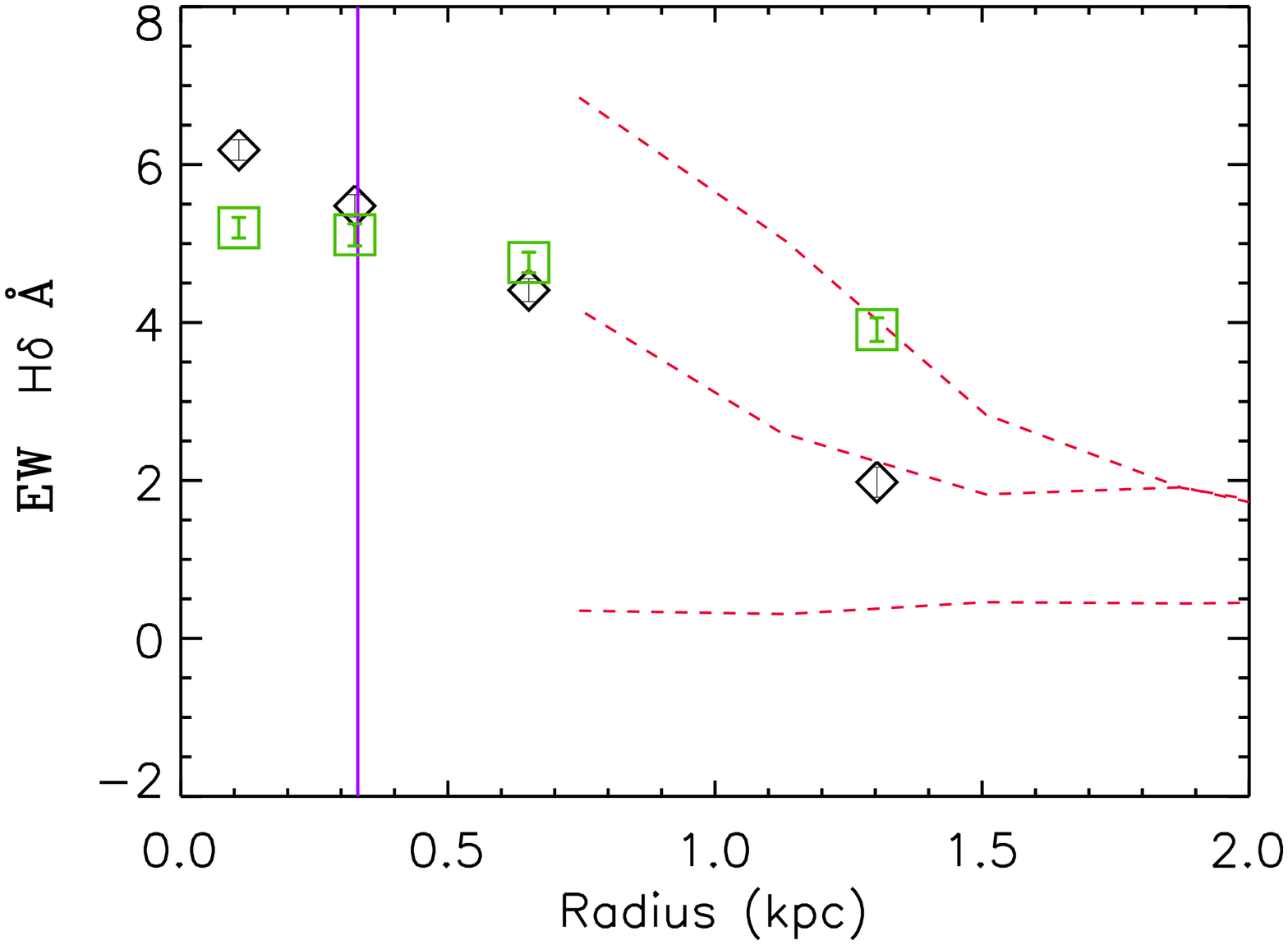}      
    \end{minipage}
    \begin{minipage}{0.95\textwidth}
      \includegraphics[width=2.9cm, angle=0, trim=0 0 0 0]{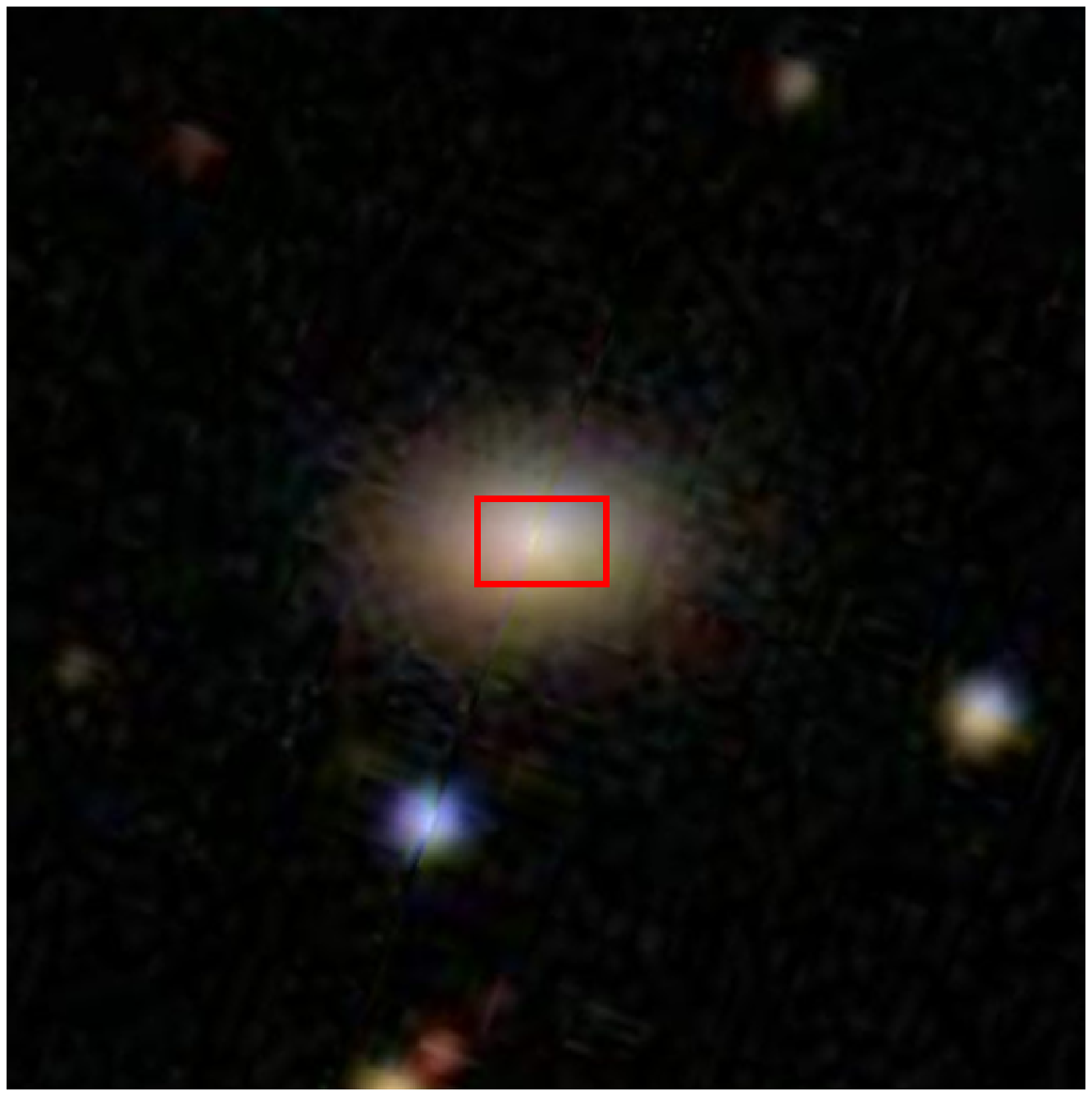}
      \includegraphics[width=5.1cm, angle=0, trim=20 35 0 0]{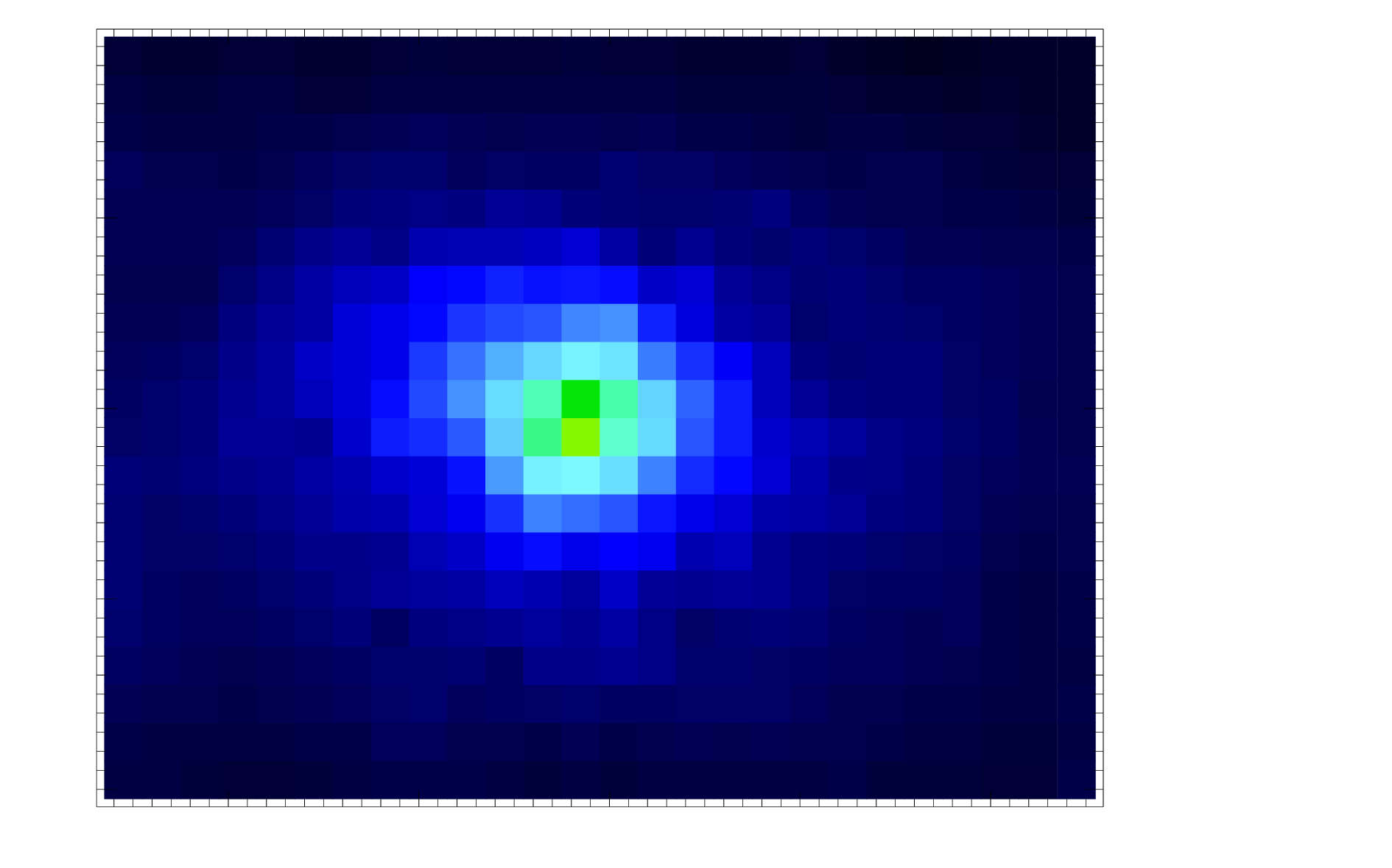}
      \includegraphics[width=4.0cm, angle=0, trim=200 35 0 0]{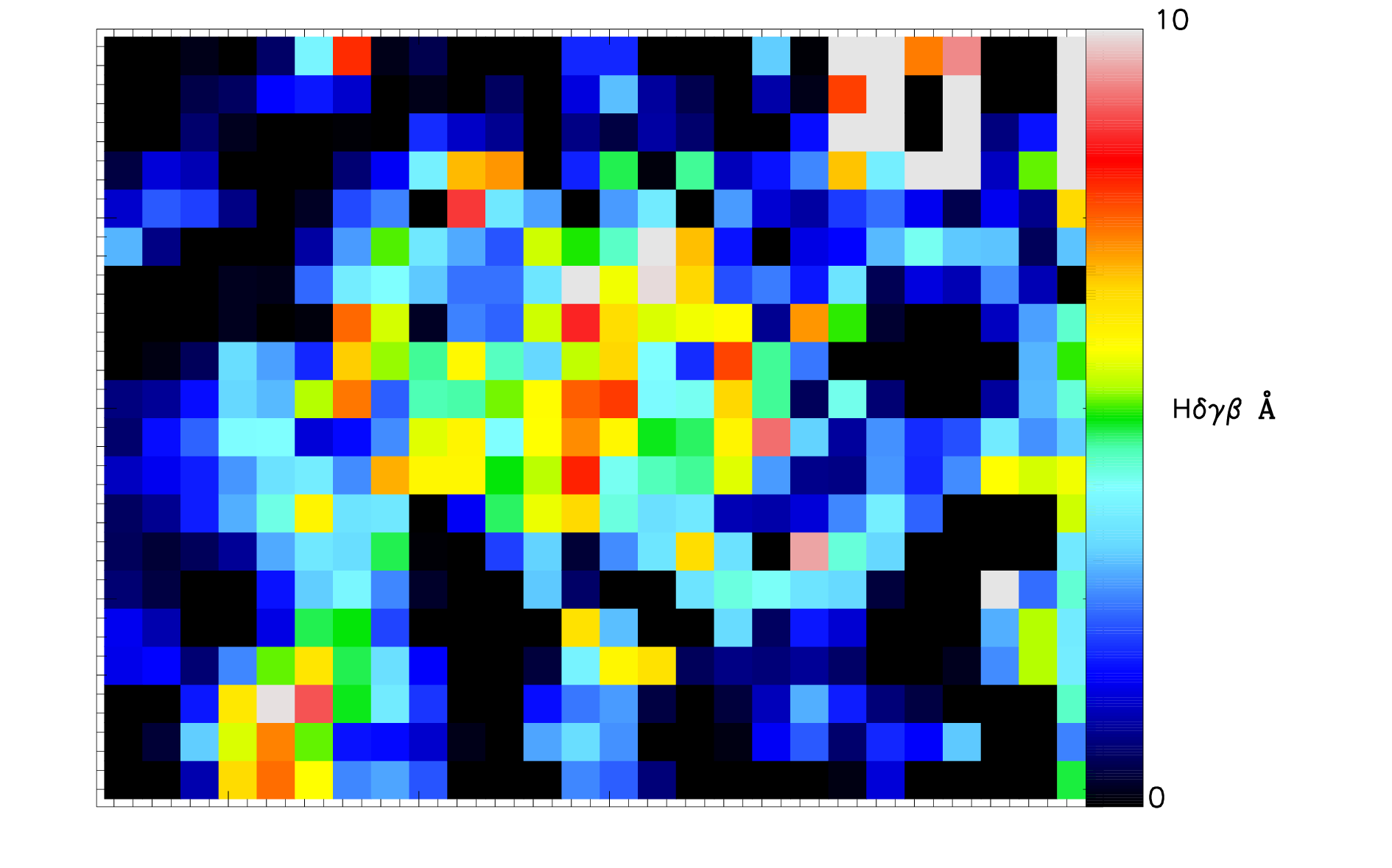}
     \includegraphics[width=3.1cm, angle=90, trim=60 0 0 100]{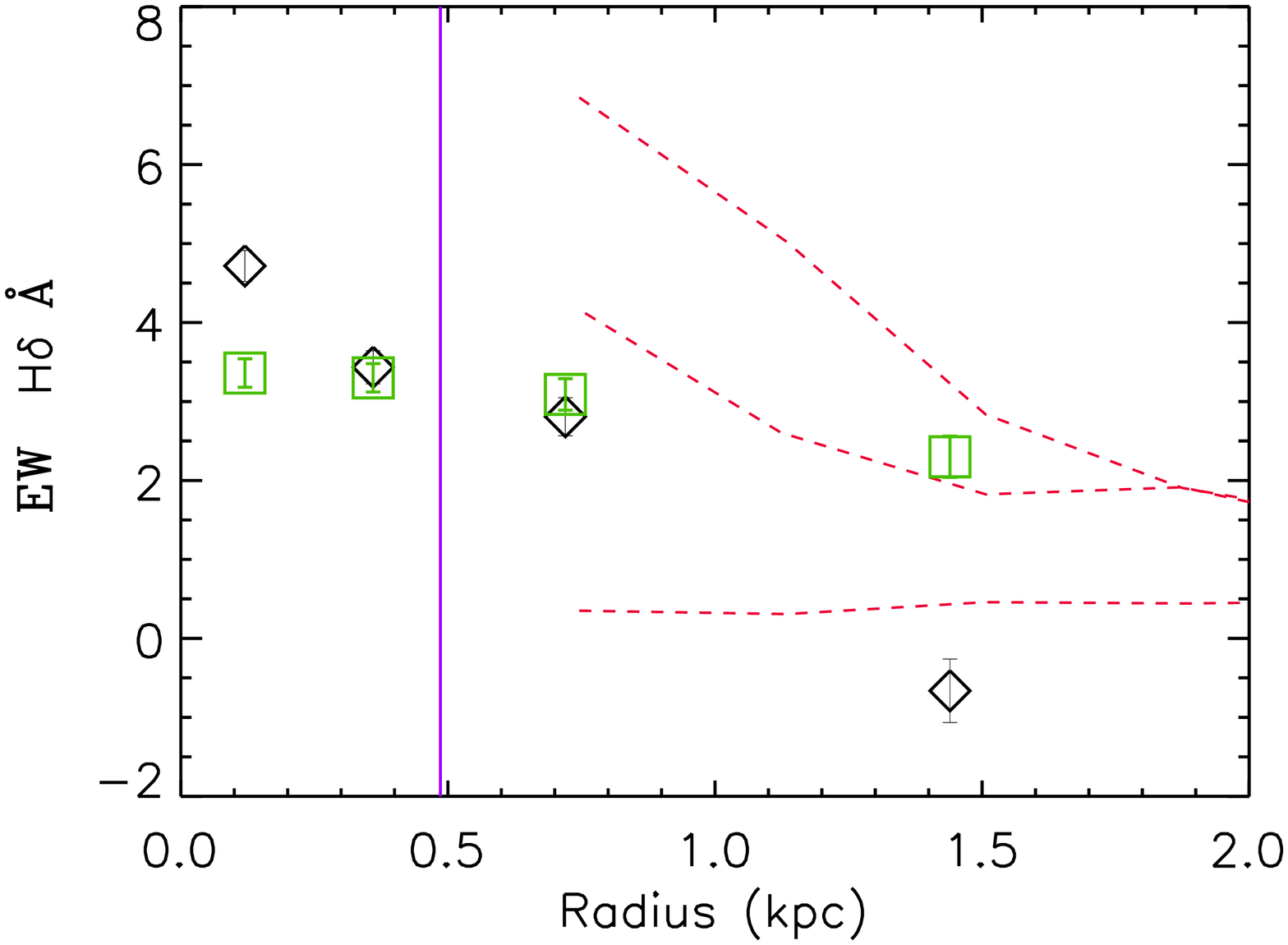}      
    \end{minipage}
    \begin{minipage}{0.95\textwidth}
      \includegraphics[width=2.9cm, angle=0, trim=0 0 0 0]{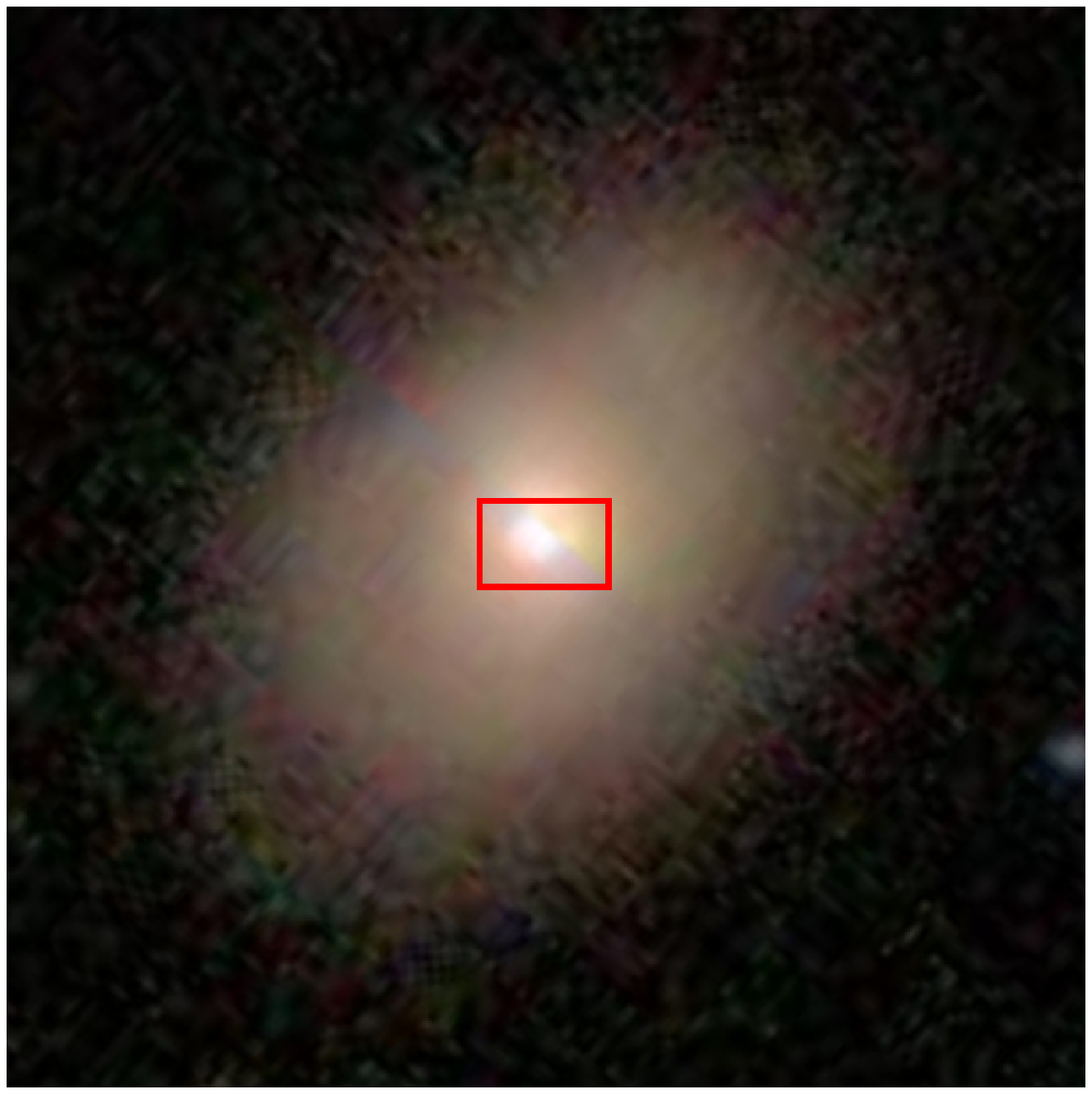}
      \includegraphics[width=5.1cm, angle=0, trim=20 35 0 0]{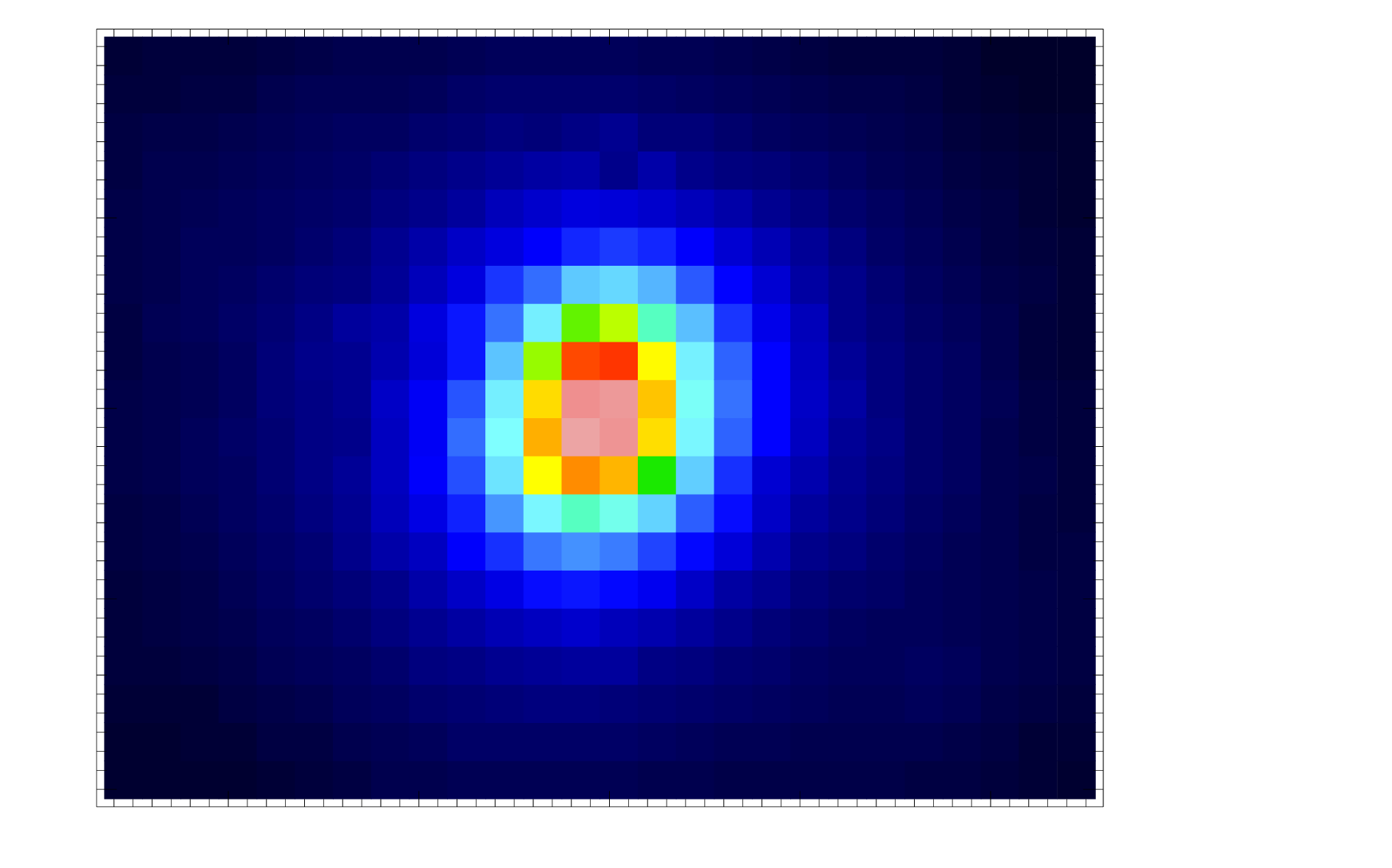}
      \includegraphics[width=4.0cm, angle=0, trim=200 35 0 0]{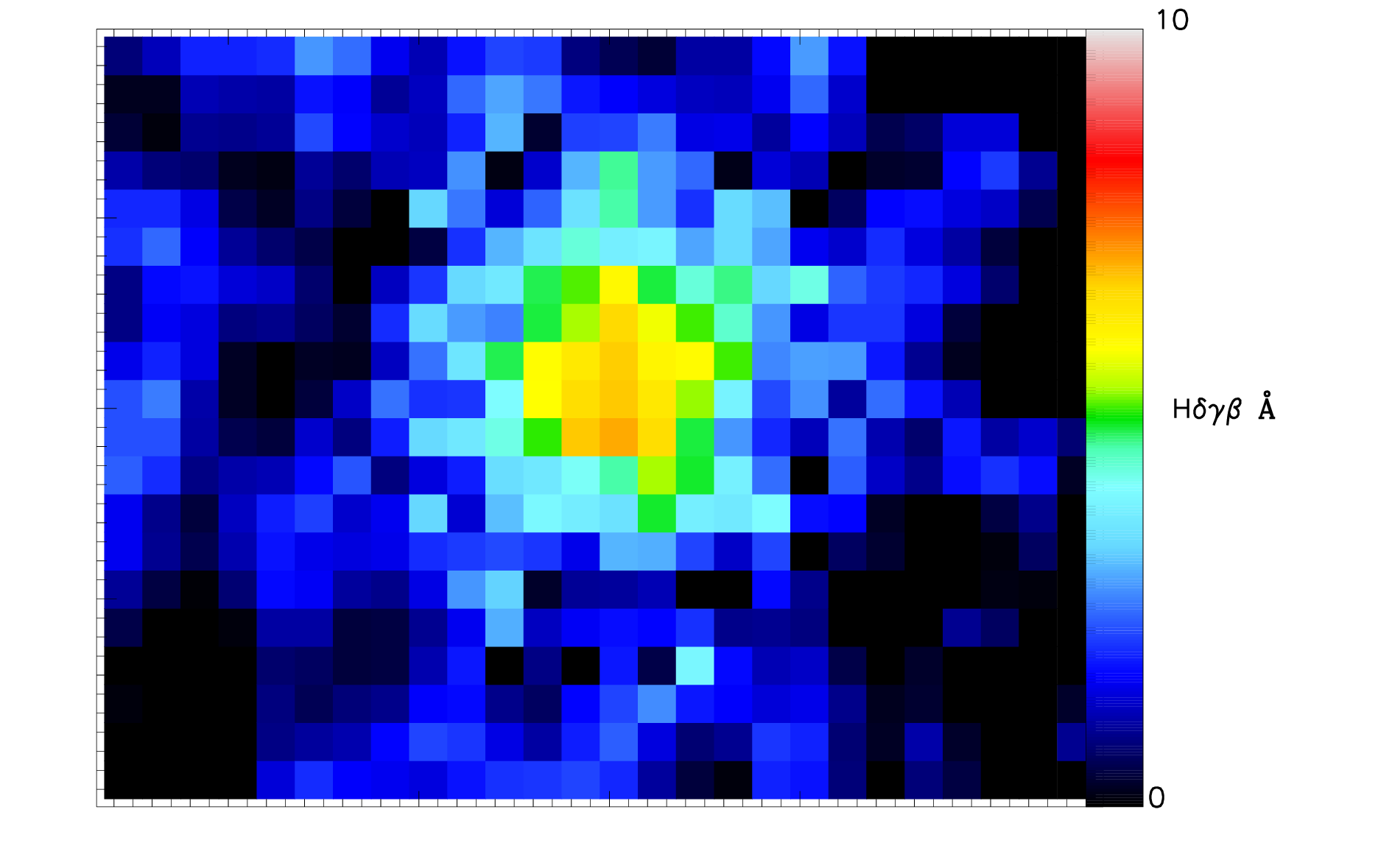}
     \includegraphics[width=3.1cm, angle=90, trim=60 0 0 100]{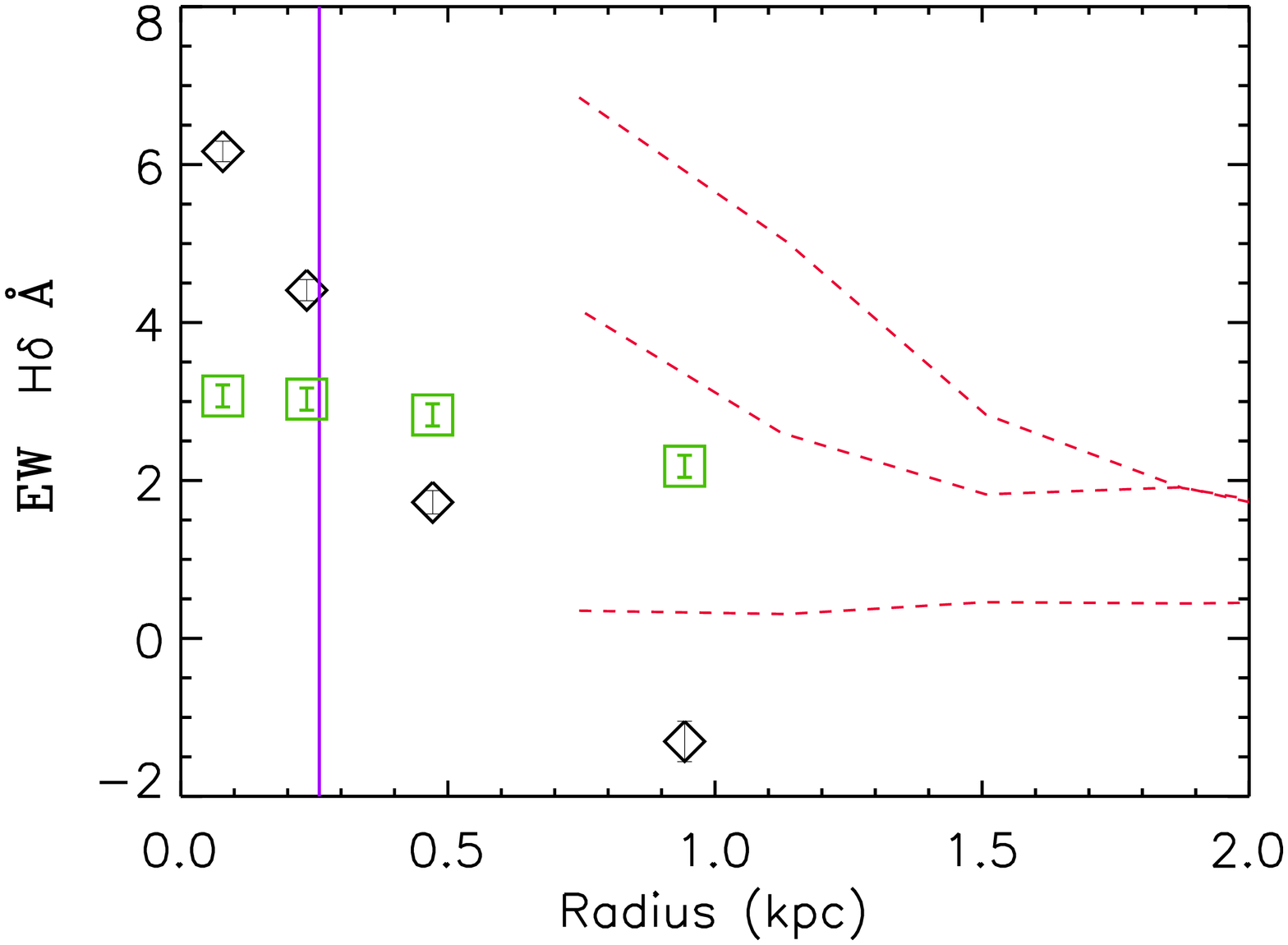}      
    \end{minipage}
\end{center}
    \caption{One object per row: SJ1613+5103; SJ2114+0032; SJ1718+3007; SJ0044-0853 ({\it top to bottom}). Columns from left to right are:
SDSS image of the target galaxy with IFU field-of-view superimposed in red; Image taken through the IFU produced by collapsing the data cube in
the wavelength direction; Balmer line equivalent width map constructed using the average strength of the H$\delta$, H$\gamma$, and H$\beta$ lines;
H$\delta$ equivalent width profiles measured from the annular spectra ({\it black diamonds}). The full width half maximum of the seeing disk is over-plotted
as a vertical blue line. The {\it green squares} are the H$\delta$ equivalent width profiles measured after the data are artificially smoothed to simulate being observed at $z=0.1$. 
The {\it red dashed} lines show the H$\delta$ profiles from the merger simulation of \citet[][see also \citet{bekki05}]{pracy05} at ages of 0.2Gyr (steepest),
0.75Gyr and 1.5Gyr (flattest) for comparison.}
    \label{fig:images}
\end{figure*}

The results are shown in the third column of Figure \ref{fig:images} which shows the Balmer line strength
over the IFU field-of-view. We have taken the average of the three Balmer lines (H$\delta$, H$\gamma$, and H$\beta$) for increased signal-to-noise but
the results are qualitatively the same using any one individual line. The Balmer equivalent width strength maps can be compared with the
galaxy continuum profile in column 2 of Figure \ref{fig:images} (which shows the IFU image constructed by collapsing the data cube in the wavelength direction)
and also to the SDSS image of each galaxy in column 1 (which has the IFU field-of-view superimposed). In all four cases the Balmer line absorption strength is
centrally concentrated, consistent with the lower luminosity E+A galaxies in the \citet{pracy12} sample.
We also measured Balmer line  equivalent width radial profiles from our annular binned spectra. The radial H$\delta$ profiles
are shown in the final column of Figure \ref{fig:images} as {\it black diamonds}. {\it All four galaxies have  centrally concentrated H$\delta$ profiles within the
central 1\,kpc}. For comparison we over-plot ({\it as red dashed lines}) the H$\delta$ profiles of a simulated merger 
from \citet[][see also \citealt{bekki05}]{pracy05} at times of 0.2\,Gyr (steepest profile), 0.75\,Gyr (middle profile) and 1.5\,Gyr (flat profile) after the
peak of the starburst.

The primary motivation for observing a `low' redshift sample of E+A galaxies was to circumvent the problems associated with beam smearing -- which can be severe \citep[e.g.][]{pracy10}. We can simulate how the radial profiles would appear if they were observed at higher redshift by artificially smoothing the data cubes to the physical scale resolution which would have been achieved by the observations for targets at that redshift.  
In this case the IFU fields do not extend over the entire galaxy so there are unavoidable edge effects in applying this artificial smoothing, nevertheless, it
demonstrates the difficulty of such observations at higher redshift. In Figure \ref{fig:images} we over-plot {\it as green squares} the radial profiles calculated from our data artificially smoothed to $z=0.1$ (i.e. similar to that of the \citet{pracy09} and \citet{swinbank12} samples) and with the same seeing as the original observations. The radial gradients in H$\delta$ are no longer observed and the profiles are essentially flat. These degraded profiles could be interpreted as galaxy-wide E+A signatures. The higher redshift studies (using the same instrument) have the advantage of covering a larger physical area, while our observations are restricted to the galaxy centre (see column  1 of Figure \ref{fig:images}).

Long slit spatially resolved observations of SJ1613+5103 (top row in Figure 3) have been performed using the FOCAS spectrograph on the Subaru telescope \citep[][]{yagi06b,goto08b}. The slit position dissects both nuclei of the interacting pair and has a position angle close to that of the long axis of our IFU observations. They report that the E+A signature is extended from the centre to at least 5\,kpc out into the `tidal plume' and there is no evidence for a H$\delta$ gradient. This conflicts with the results displayed in Figure 3 where there is a clear H$\delta$ and Balmer line gradient; albeit the shallowest one in our sample. There is also disagreement in the absolute strength of the H$\delta$ equivalent width -- although this can depend on the precise index and continuum bands used as well as the size and placement of the aperture. 
\citet{yagi06b} measure the central H$\delta$ equivalent width to be $\sim$6--7\,\AA\, while  \citet{goto08b}, using the same definitions, find H$\delta$ equivalent widths of $\sim$7-10\AA. In this work we find the central H$\delta$ equivalent width, using the H$\delta_{\rm A}$ index definition, to be only $\sim$5\AA\, (see right hand column of Figure 3). Although this should not be directly compared since, as mentioned above, we have placed our values on the Lick system. We note that in general our measurements of equivalent widths in the central aperture are in good agreement with the published SDSS values (on the same system). 
The radial profile for the H$\delta$ line measured by \citet{goto08b} is uniformly strong (with a slight negative gradient) but their profiles for H$\gamma$ and H$\beta$ have positive gradients and  weak central absorption strengths ($\sim$2--3\AA). This is inconsistent with the results presented here as well as with the SDSS catalog values which have H$\beta$ and H$\gamma$ equivalent widths of $\sim$6\AA. Both \citet{goto08b} and \citet{yagi06b} find a gradient in the D4000 break consistent with a younger stellar population in the galaxy centre. They conclude this trend in D4000 combined with the lack of a H$\delta$ gradient cannot be fully explained by existing theoretical models \citep{yagi06b}.  This galaxy is the highest redshift object in our sample and the data is of moderate signal-to-noise and resolving these inconsistencies with certainty will likely require higher quality observations.

\begin{figure*}
  \begin{center}
    \begin{minipage}{0.95\textwidth}
      \includegraphics[width=5.2cm, angle=0, trim=0 0 0 0]{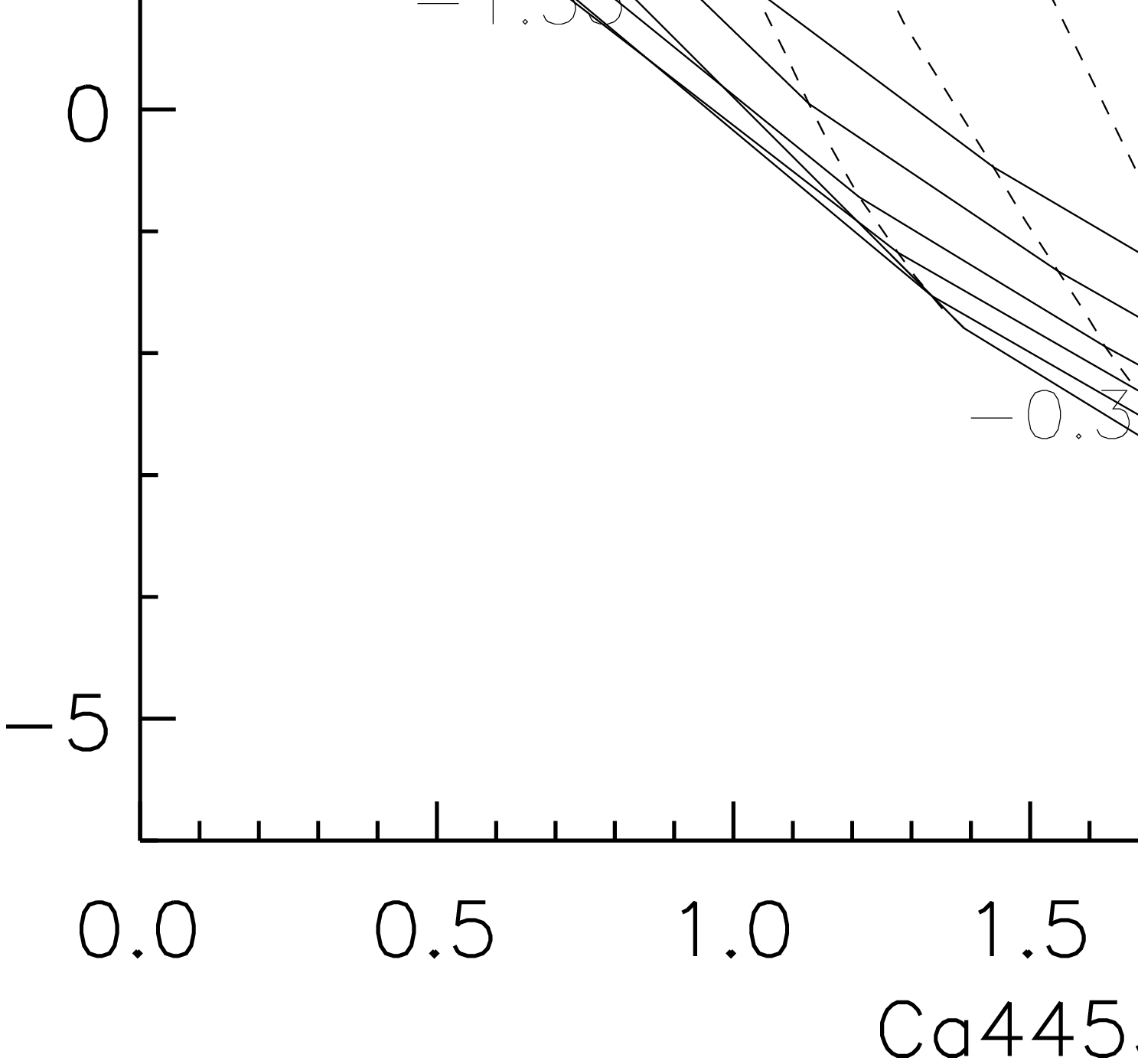}     
      \includegraphics[width=5.2cm, angle=0, trim=0 0 0 0]{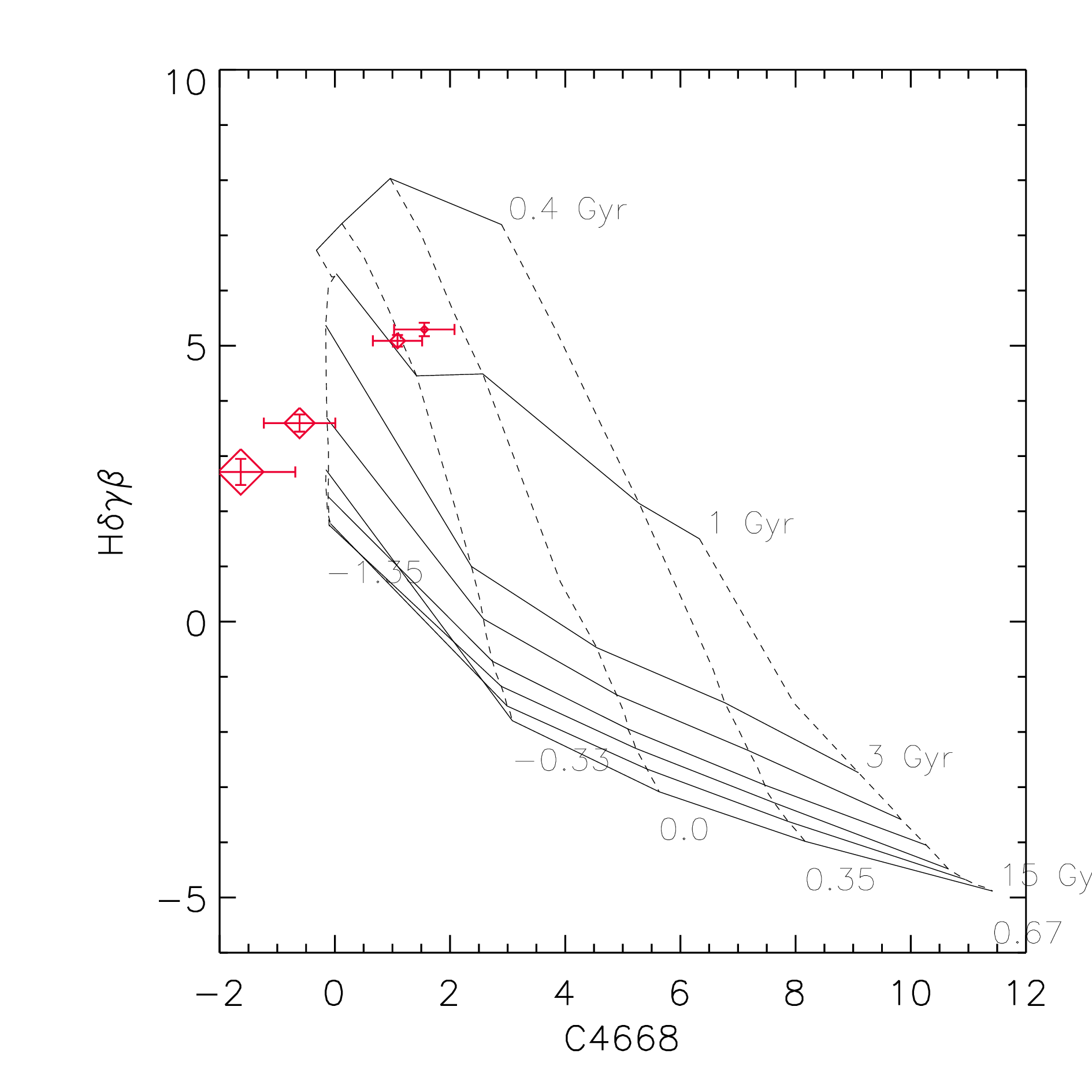}    
      \includegraphics[width=5.2cm, angle=0, trim=0 0 0 0]{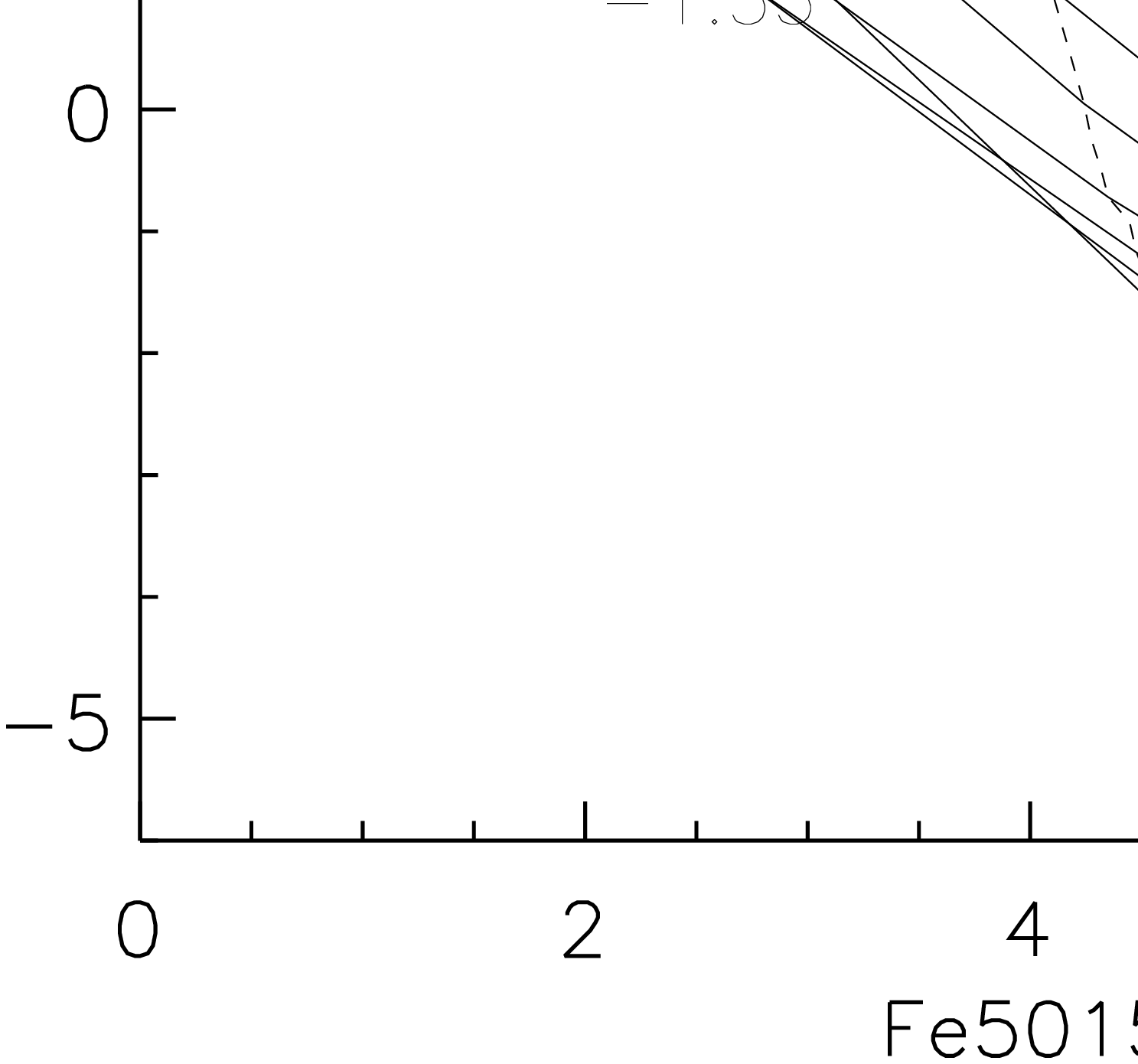}      
    \end{minipage}
    \begin{minipage}{0.95\textwidth}
      \includegraphics[width=5.2cm, angle=0, trim=0 0 0 0]{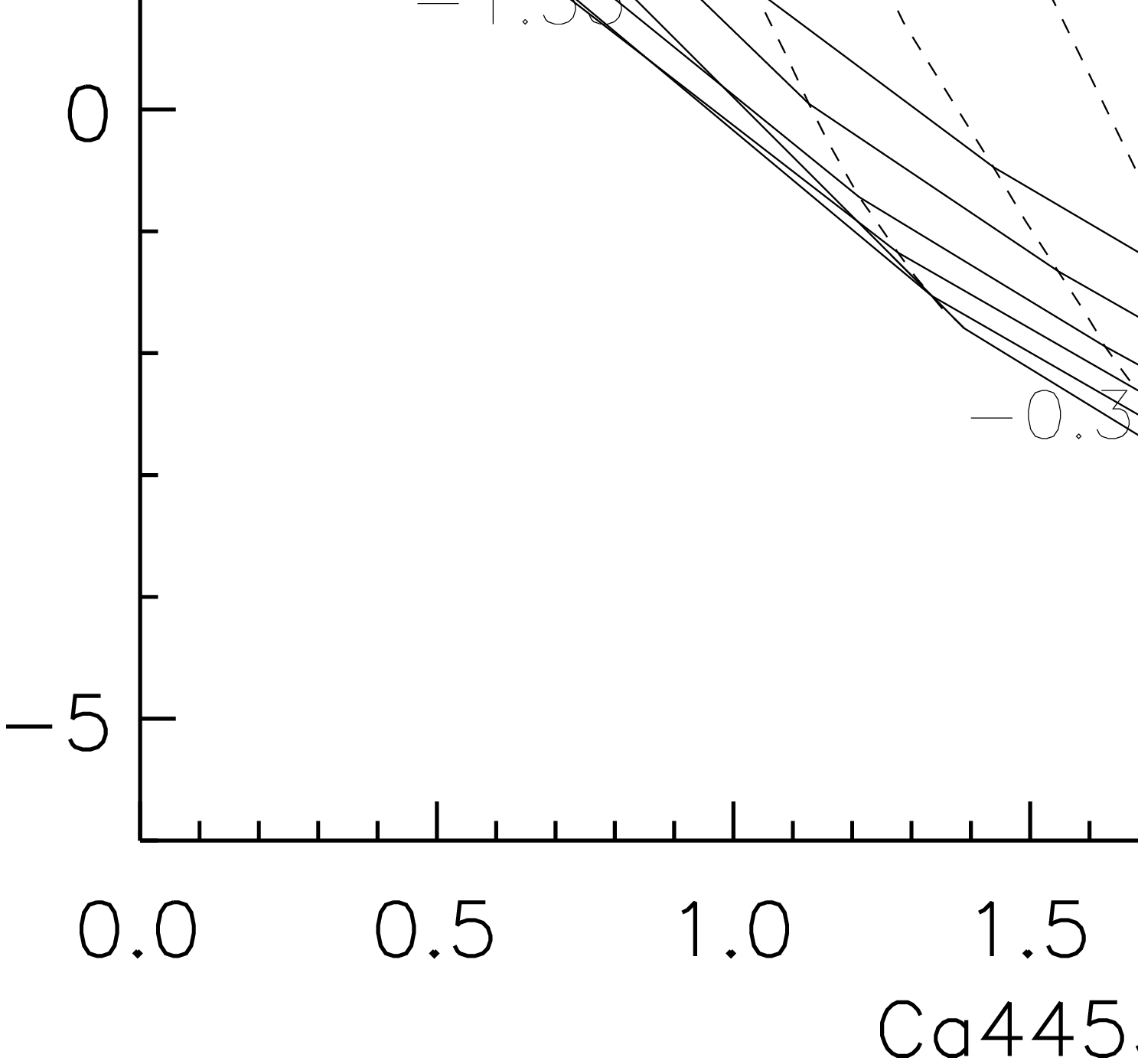}     
      \includegraphics[width=5.2cm, angle=0, trim=0 0 0 0]{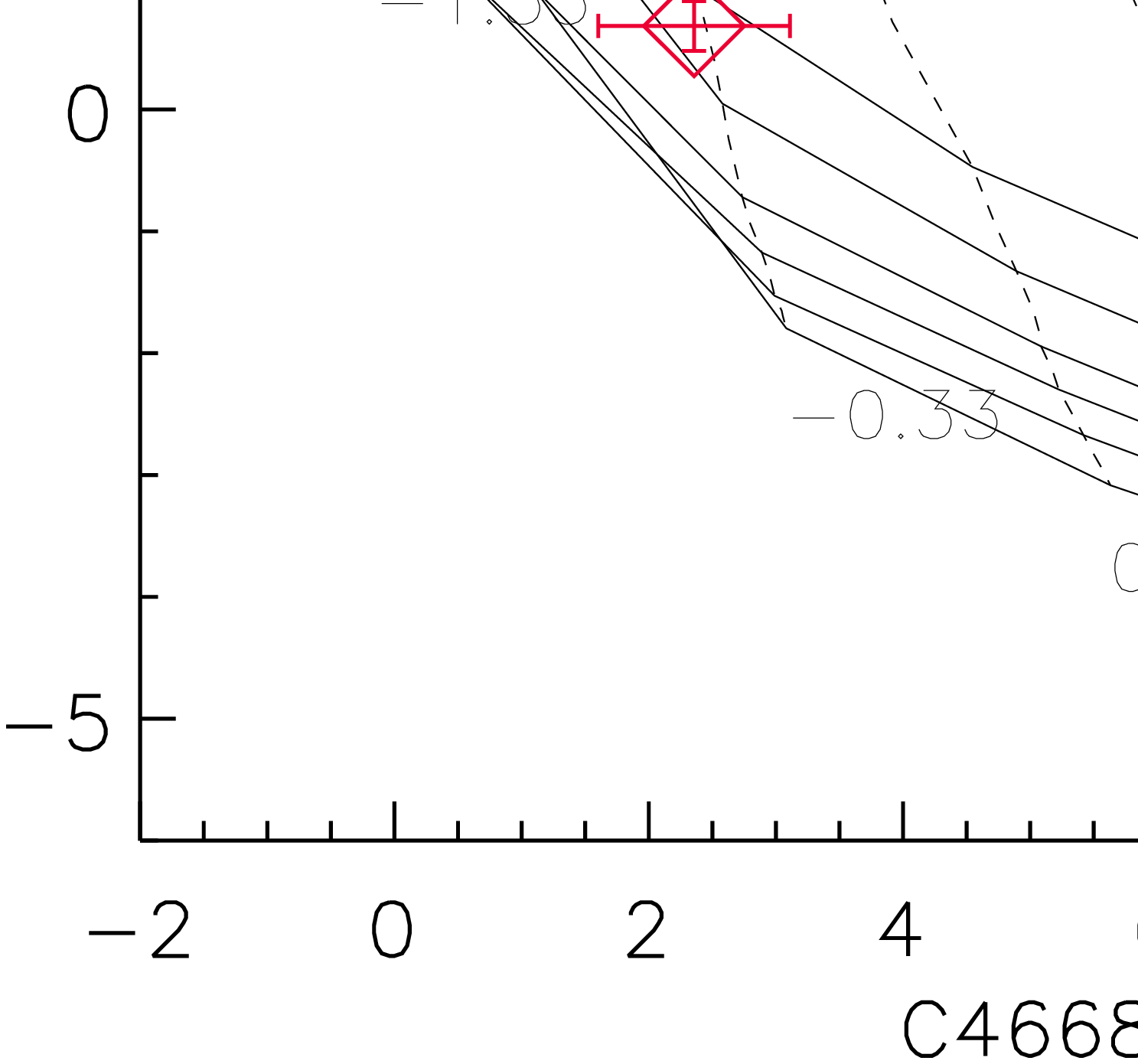}    
      \includegraphics[width=5.2cm, angle=0, trim=0 0 0 0]{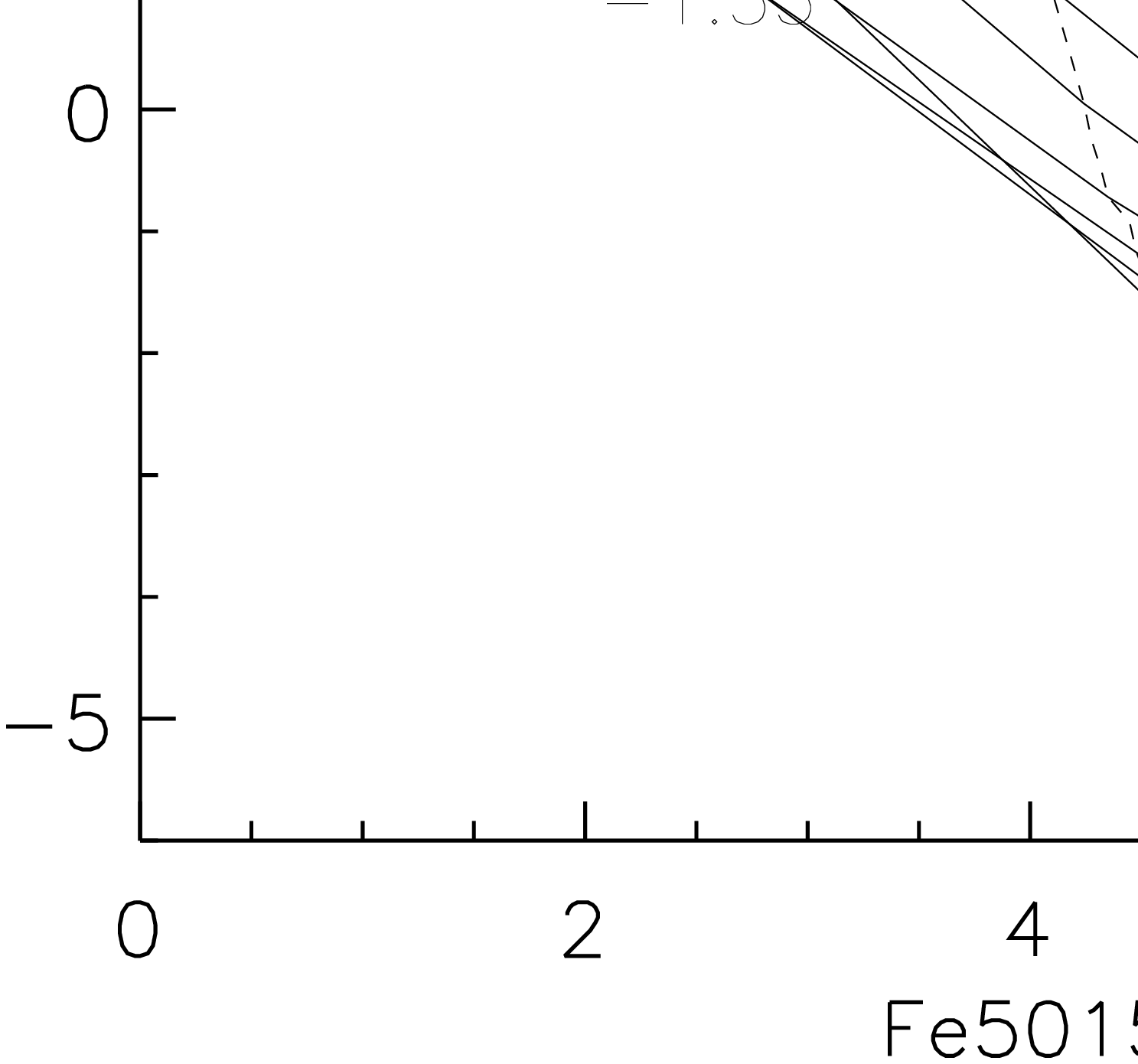}      
    \end{minipage}
    \begin{minipage}{0.95\textwidth}
      \includegraphics[width=5.2cm, angle=0, trim=0 0 0 0]{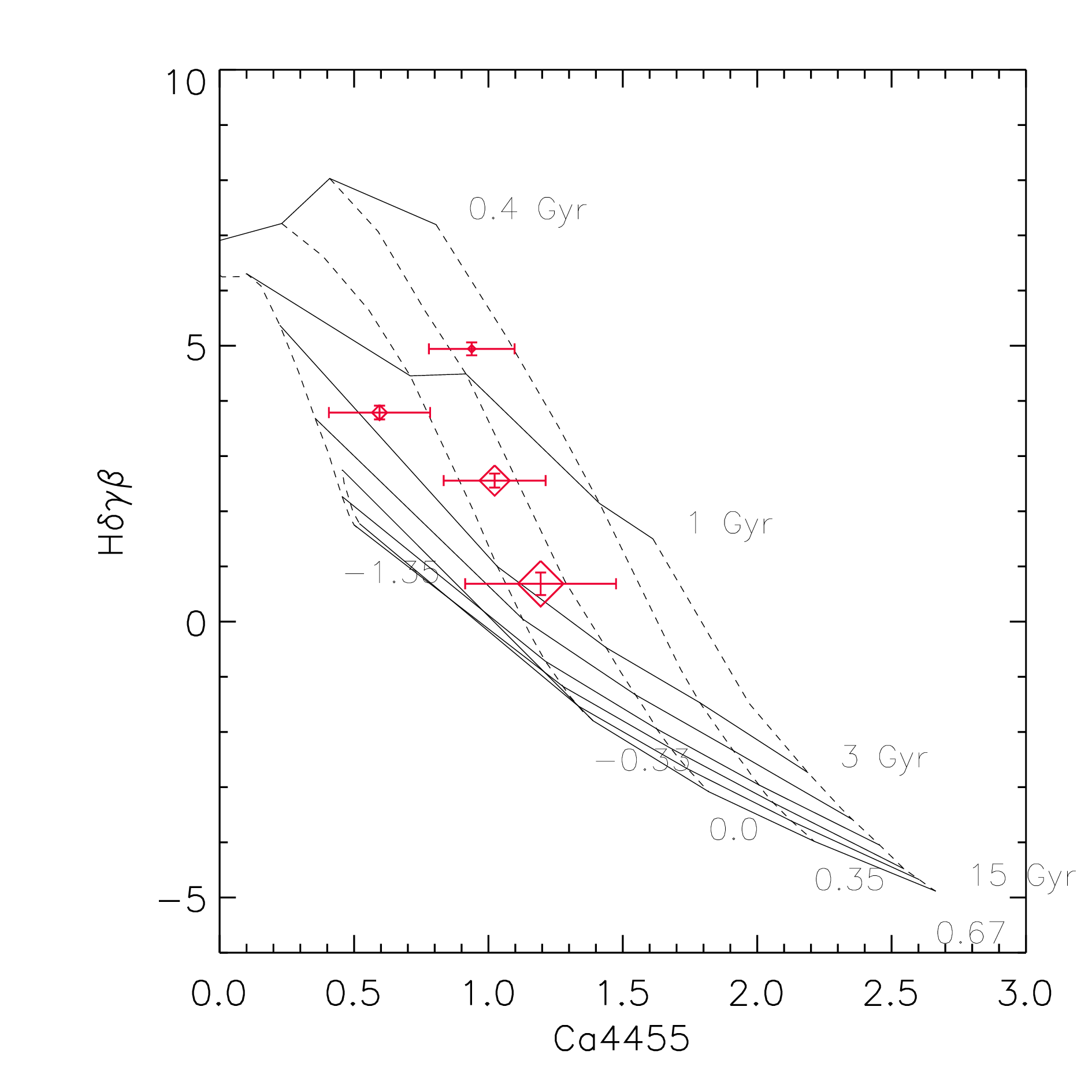}     
      \includegraphics[width=5.2cm, angle=0, trim=0 0 0 0]{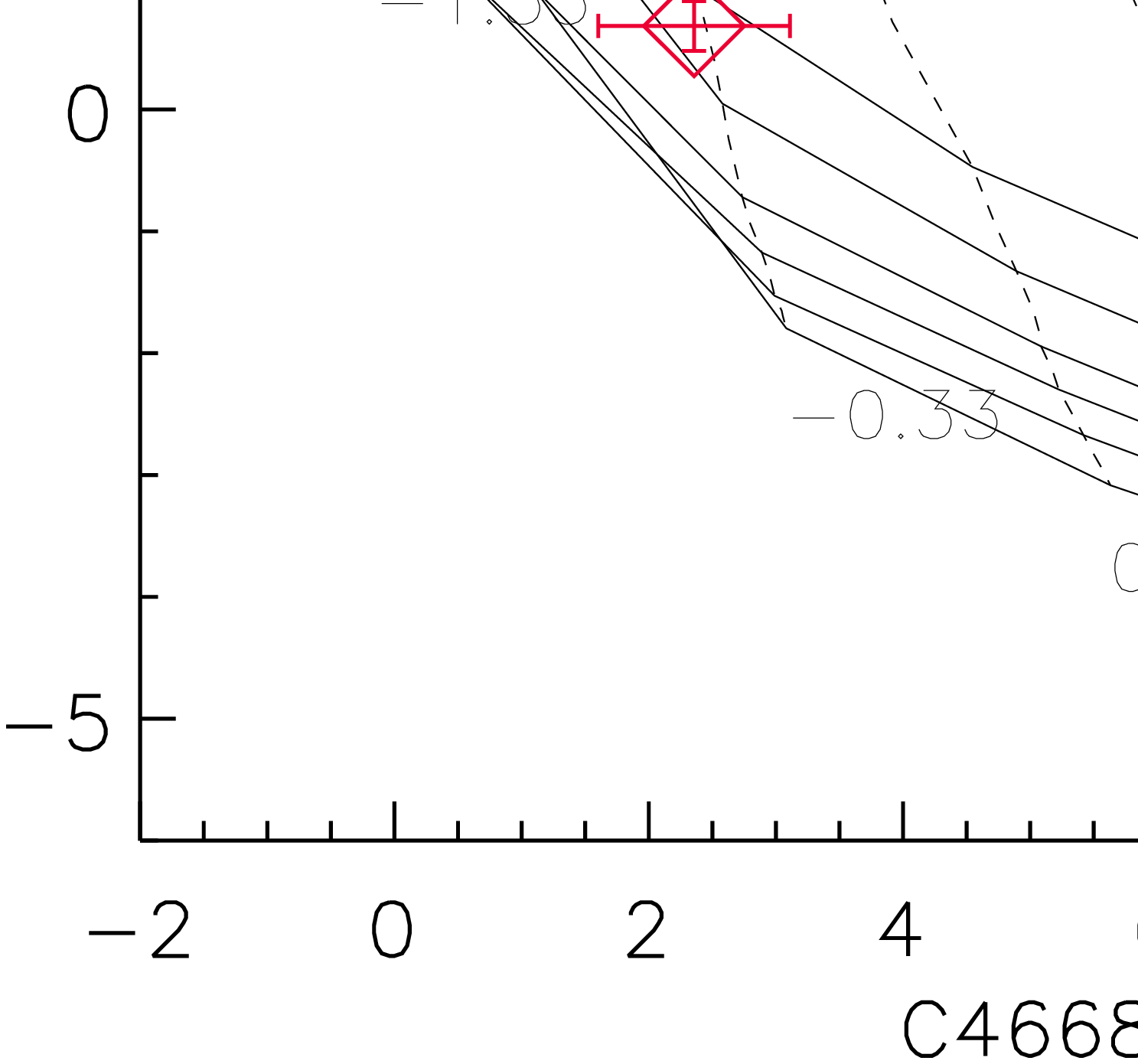}    
      \includegraphics[width=5.2cm, angle=0, trim=0 0 0 0]{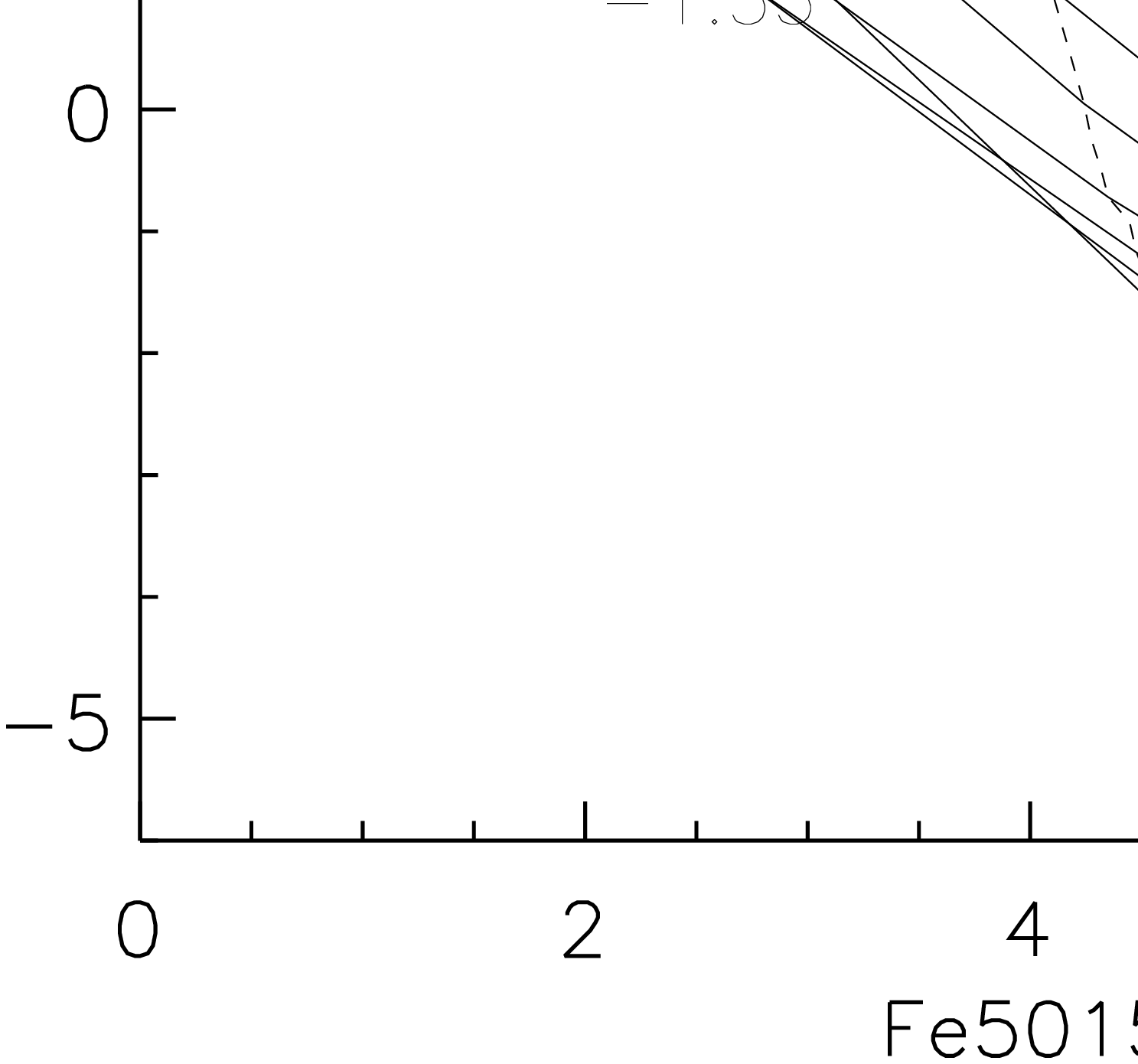}      
    \end{minipage}
    \begin{minipage}{0.95\textwidth}
      \includegraphics[width=5.2cm, angle=0, trim=0 0 0 0]{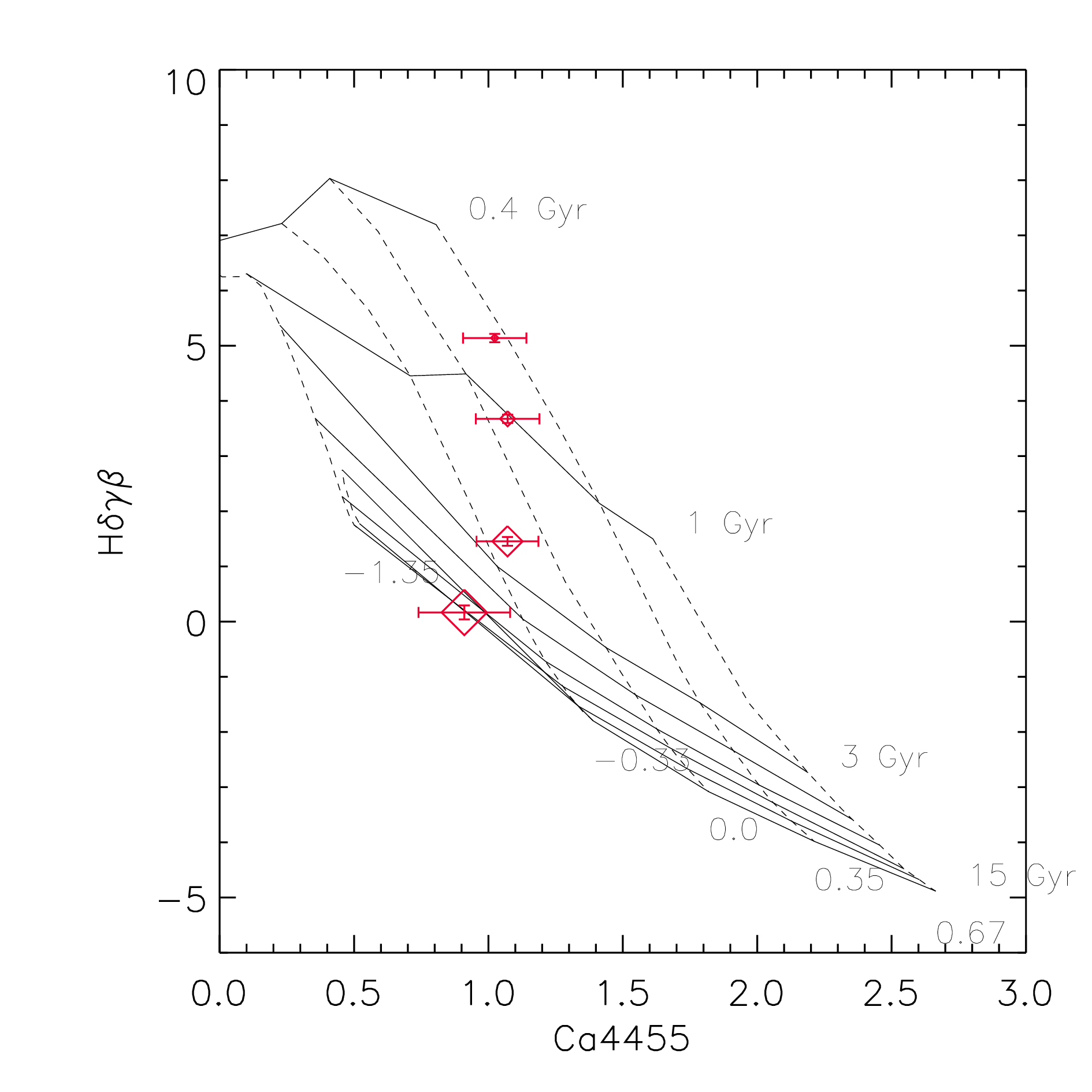}     
      \includegraphics[width=5.2cm, angle=0, trim=0 0 0 0]{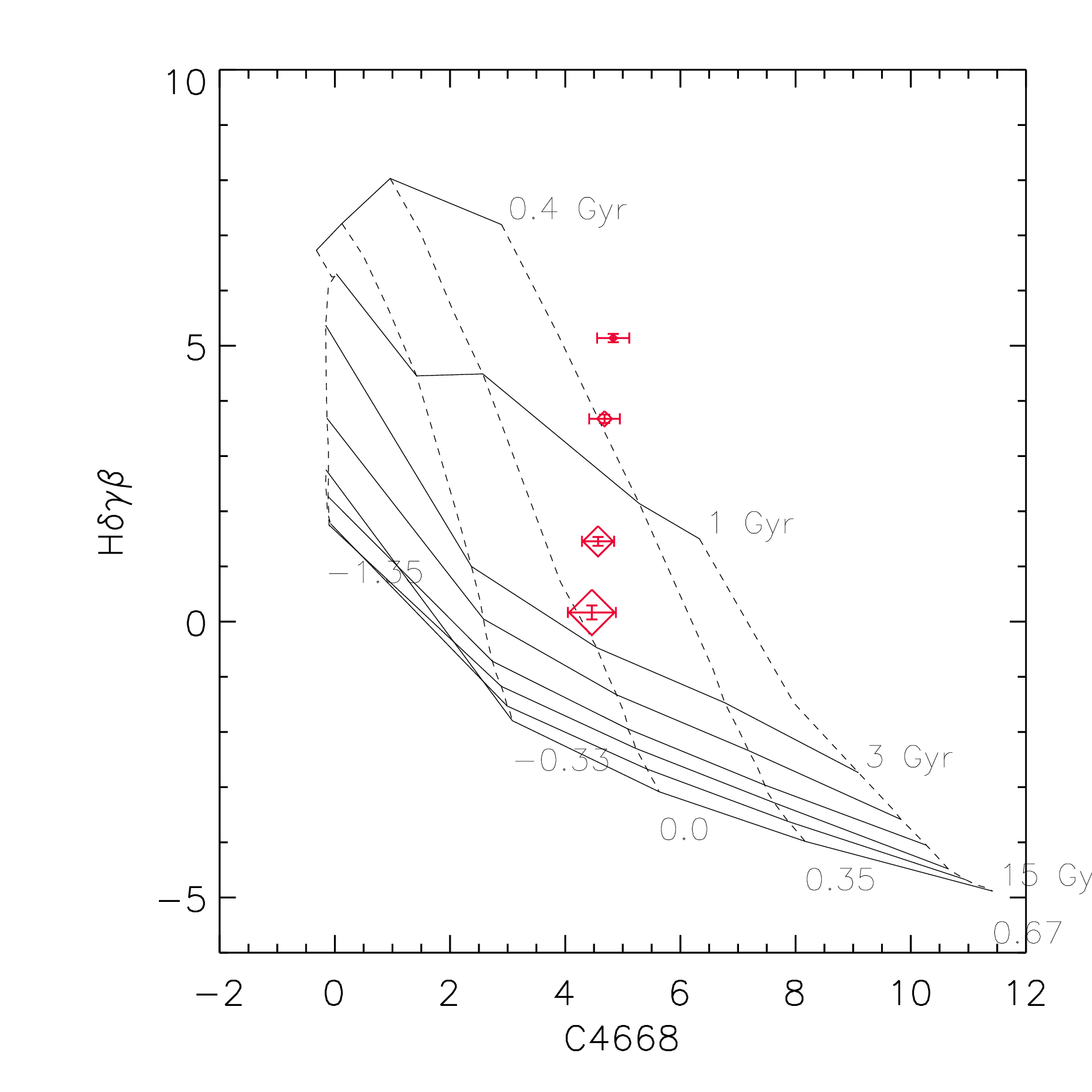}    
      \includegraphics[width=5.2cm, angle=0, trim=0 0 0 0]{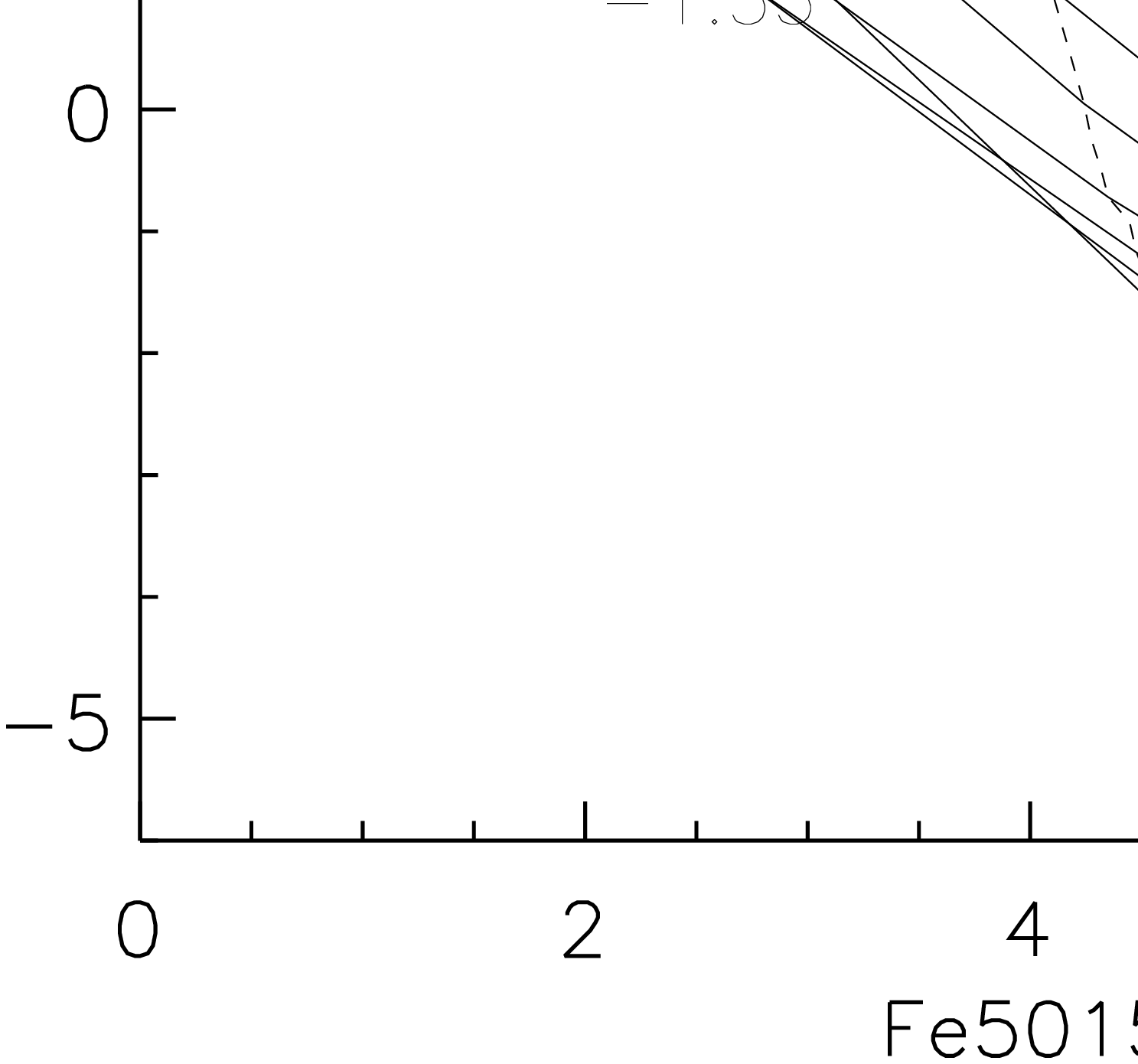}      
    \end{minipage}
\end{center}
    \caption{One object per row with the same ordering as Figure \ref{fig:images}. Age--metallicity diagnostic diagrams using the averaged Balmer line equivalent width as the age
sensitive index plotted against the metal sensitive indices Ca4455 (left), C4668 (middle) and Fe5015 (right). The over-plotted age-metallicity
grids are from the single age stellar population models of \citet{thomas03,thomas04} and assume solar abundance ratios. The red diamonds are the index values
measured from the annular spectra where increasing symbol size corresponds to increasing galacto-centric radius. Note the annular spectra are derived from the angular
size bins so correspond to different physical sizes for each galaxy. All four objects have significant
age gradients.}
    \label{fig:agemet}
\end{figure*}

That the Balmer line gradients in Figure \ref{fig:images} correspond to age gradients is demonstrated in Figure \ref{fig:agemet} where we have plotted
a series of age-metallicity diagnostic diagrams. We used the combined H$\delta$, H$\gamma$, and H$\beta$ line equivalent width for the age sensitive
index plotted against several of the standard metallicity indices within our wavelength coverage. The index values measured from the annular binned spectra are 
over-plotted as {\it red diamonds} where the symbol size is used to represent the galacto-centric radius of the annular bin -- in the sense that larger symbol sizes
represent larger radii. Over-plotted are age-metallicity grids derived from the stellar population models of \citet{thomas03,thomas04} using solar abundance ratios.
For all four galaxies there is a clear trend of increasing age with increasing radius, from $\lesssim$ 1\,Gyr in the galaxy centres to 
a few Gyr in the outermost annuli. There is more scatter and some systematic differences between indices in metallicity and no strong radial trends. There is a hint of a
negative metallicity gradient in SJ0044-0853 ({\it fourth row}) with higher metallicity values in the galaxy centre.

\subsection{Kinematics}
\begin{figure*}
  \begin{center}
    \begin{minipage}{0.95\textwidth}
      \includegraphics[width=8.4cm, angle=0, trim=0 0 0 0]{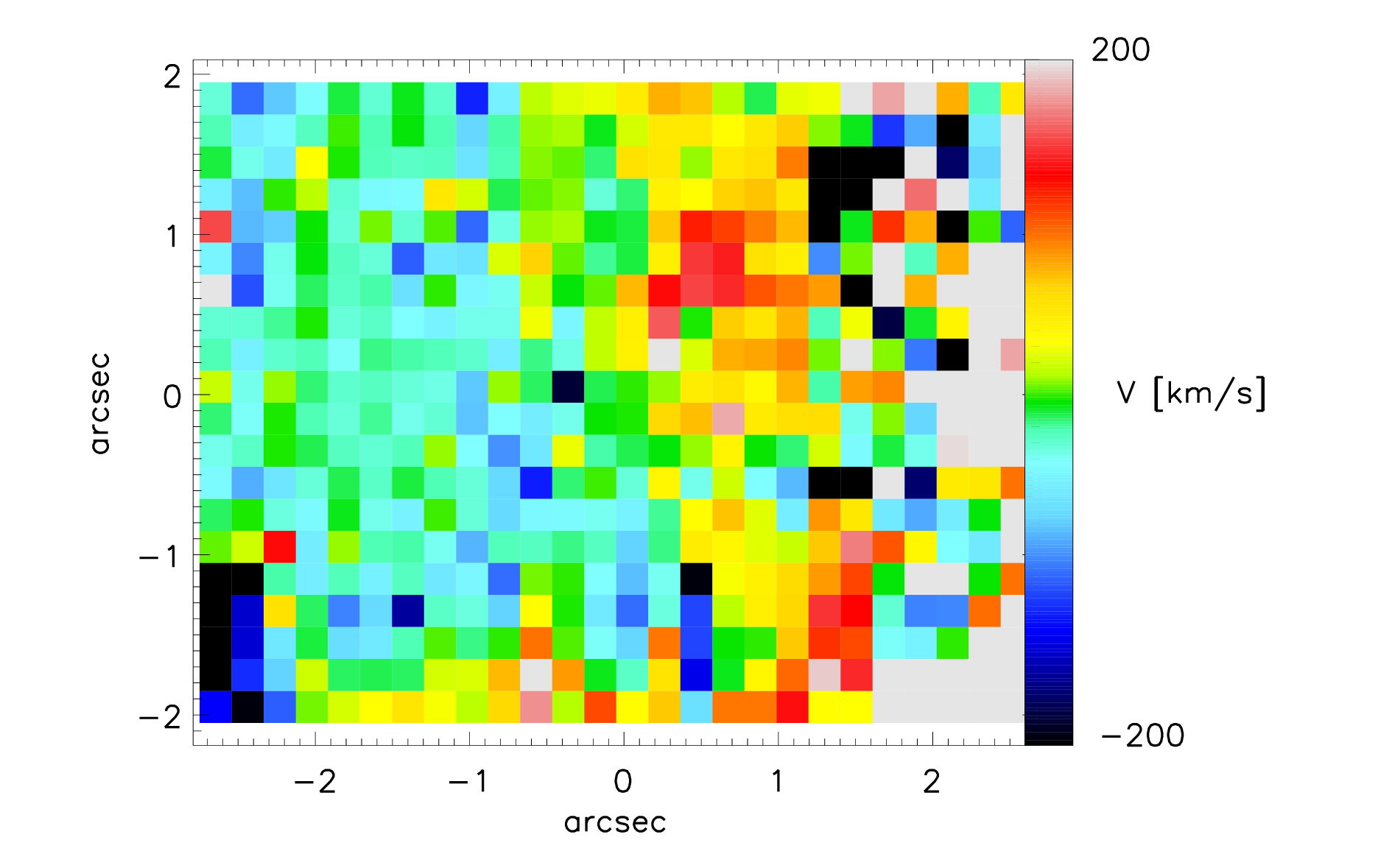}     
      \includegraphics[width=5.4cm, angle=90, trim=0 0 0 0]{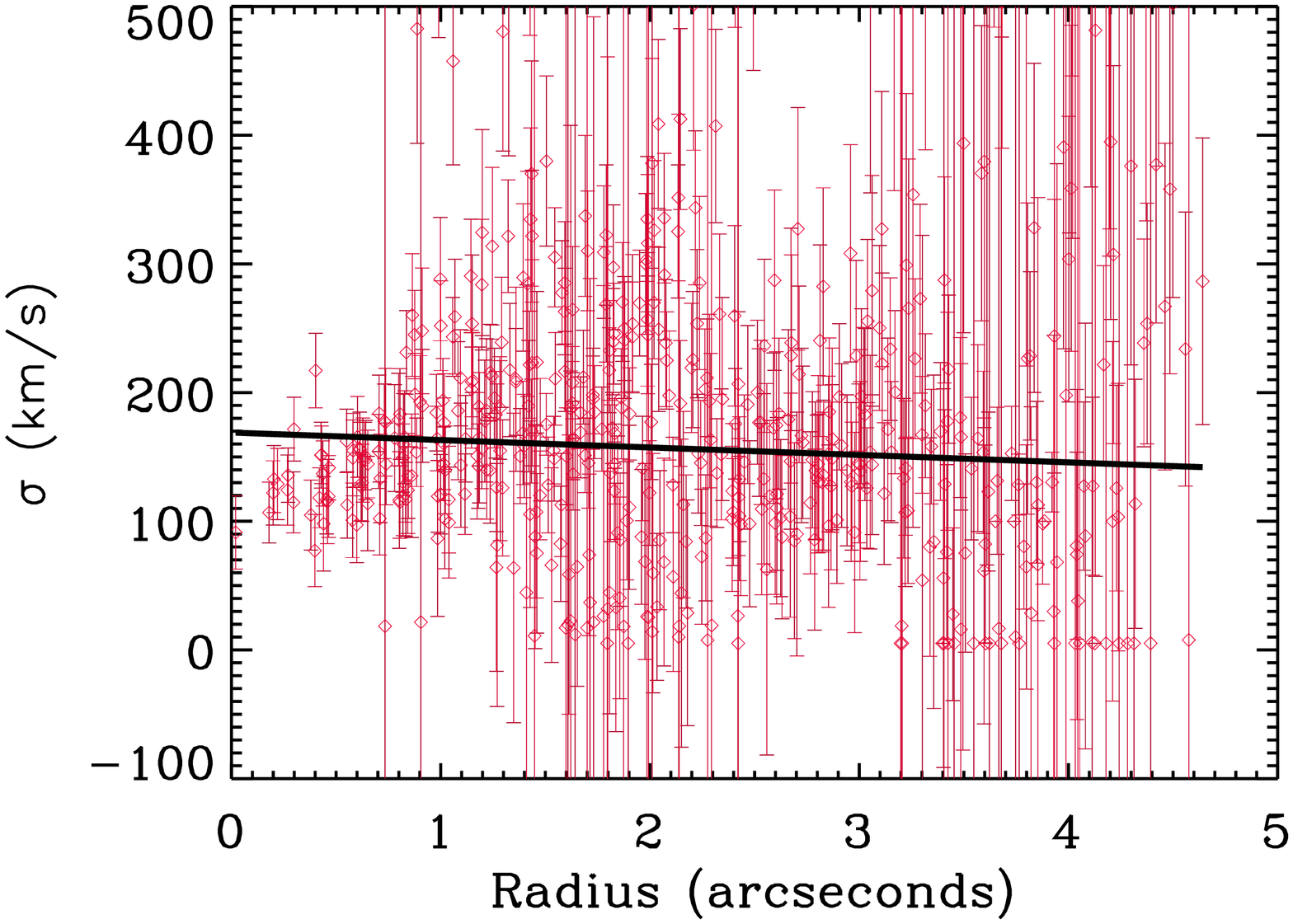}         
    \end{minipage}
    \begin{minipage}{0.95\textwidth}
      \includegraphics[width=8.4cm, angle=0, trim=0 0 0 0]{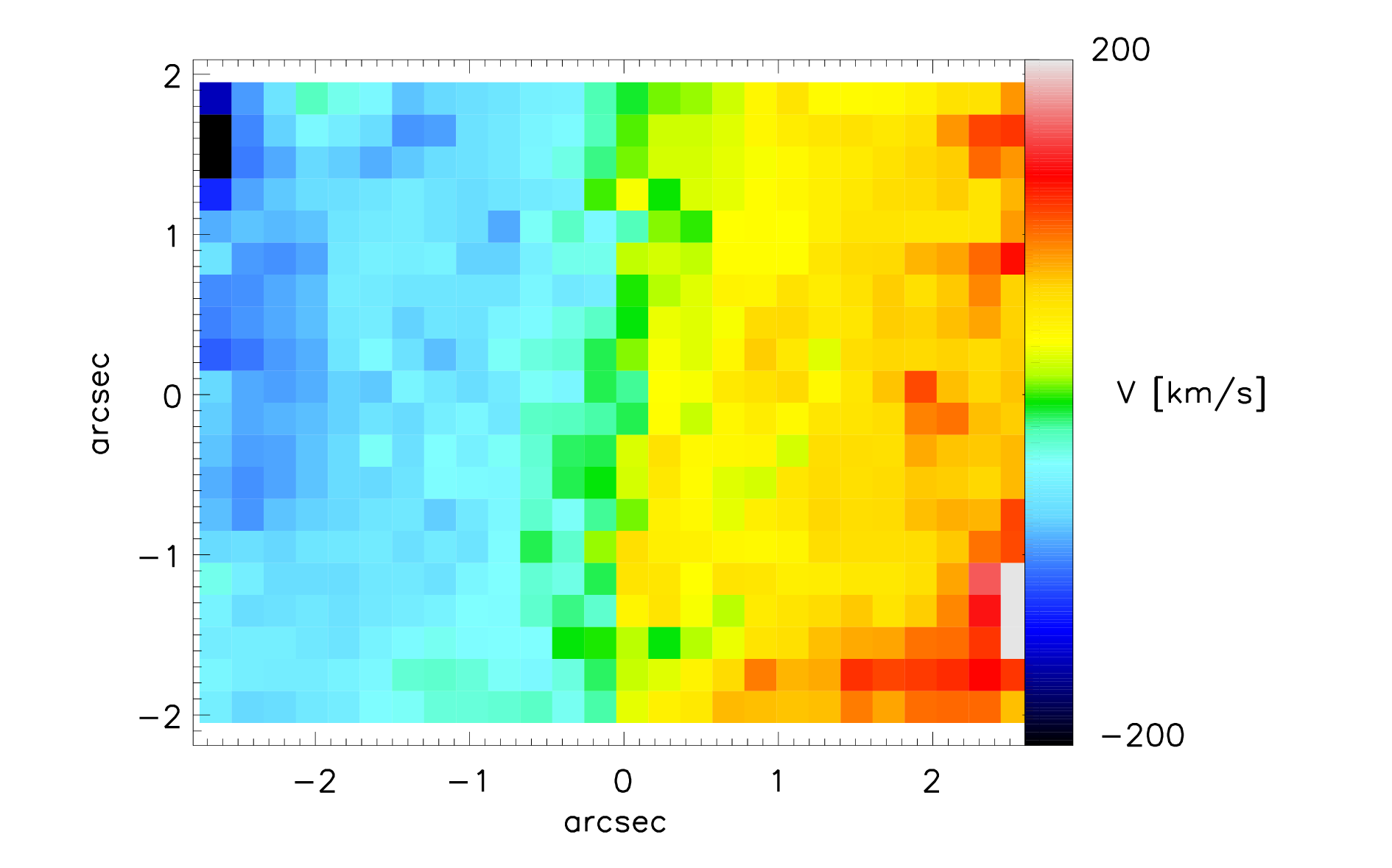}     
      \includegraphics[width=5.4cm, angle=90, trim=0 0 0 0]{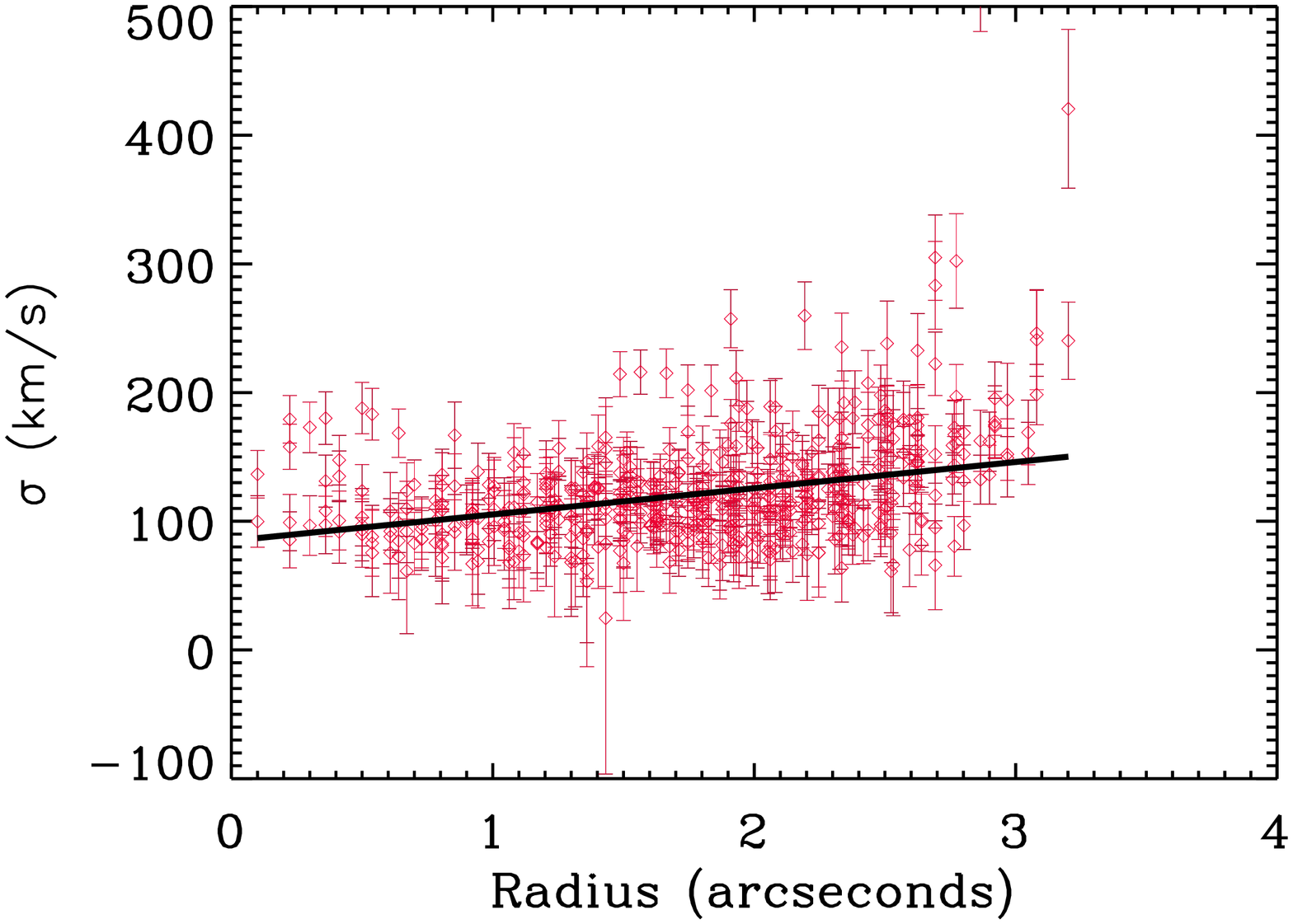}       
    \end{minipage}
    \begin{minipage}{0.95\textwidth}
      \includegraphics[width=8.4cm, angle=0, trim=0 0 0 0]{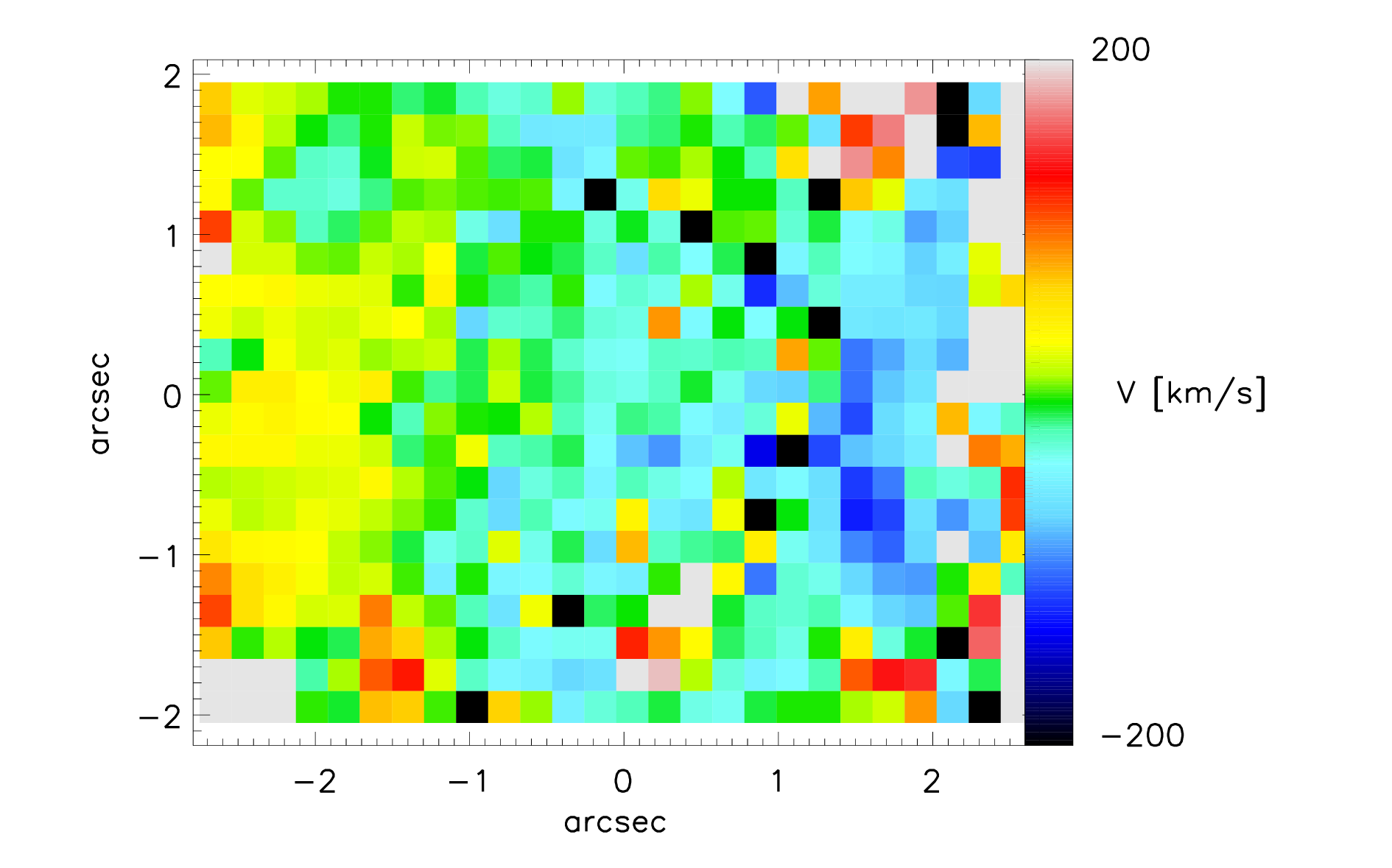}     
      \includegraphics[width=5.4cm, angle=90, trim=0 0 0 0]{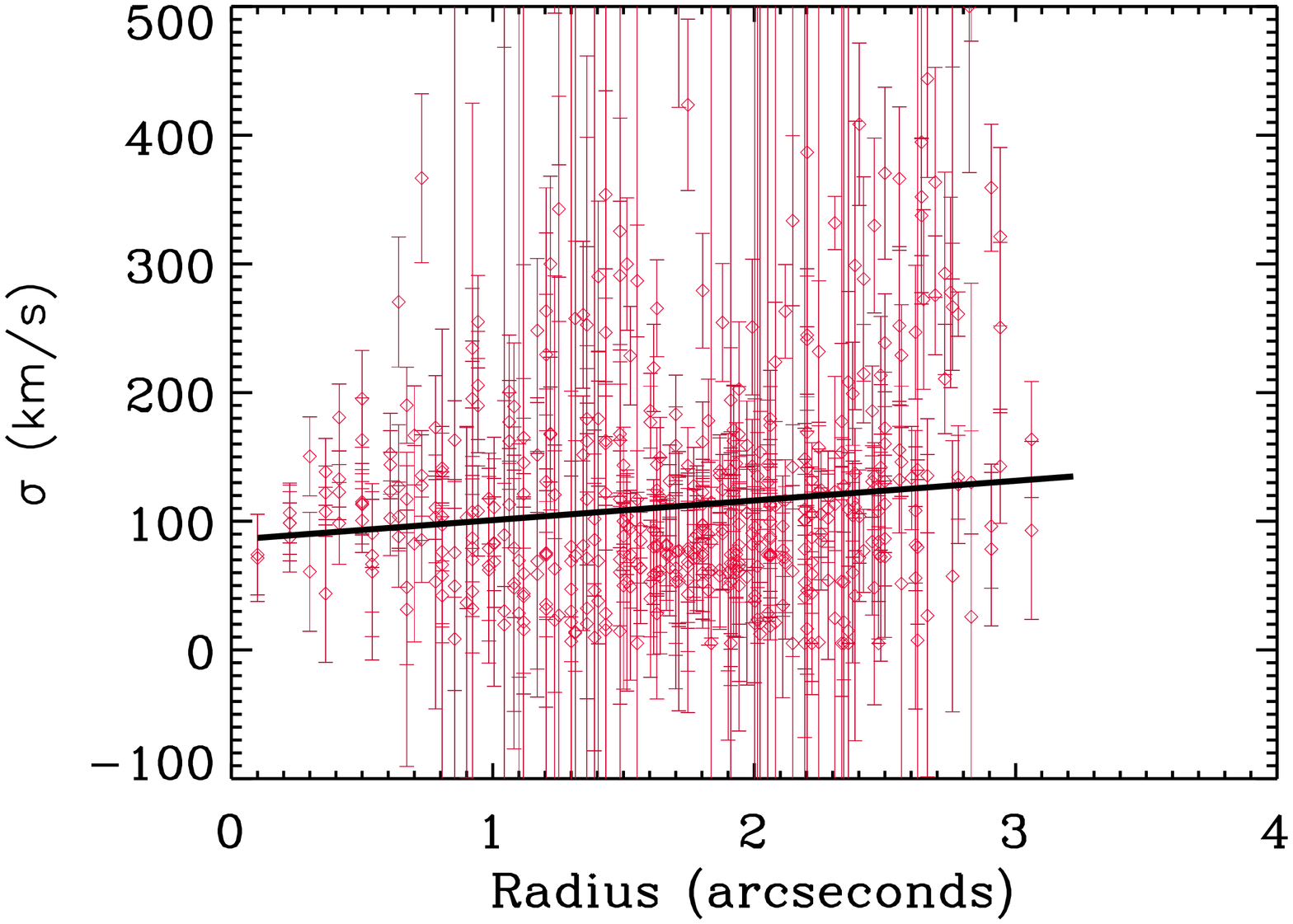}         
    \end{minipage}
    \begin{minipage}{0.95\textwidth}
      \includegraphics[width=8.4cm, angle=0, trim=0 0 0 0]{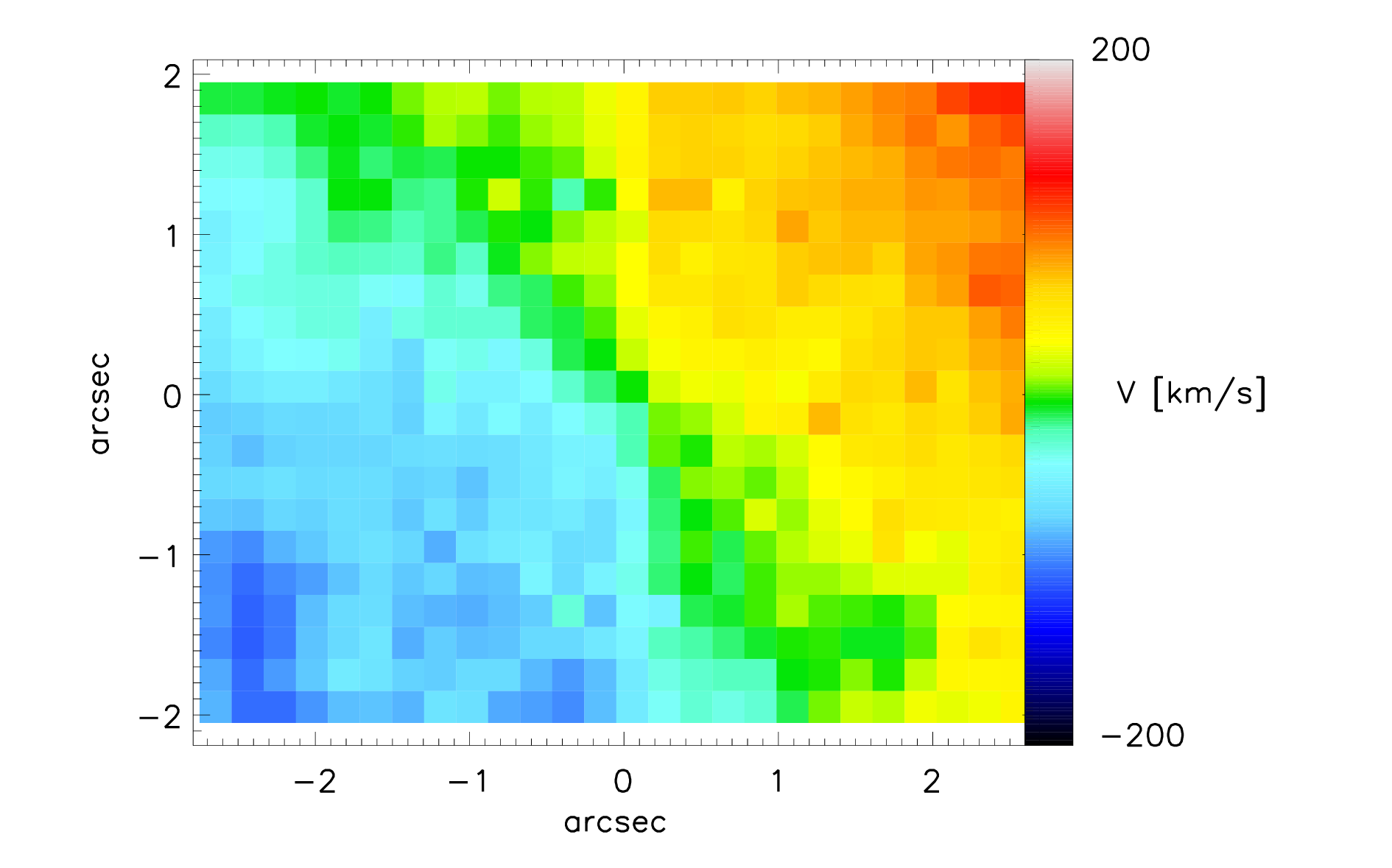}     
      \includegraphics[width=5.4cm, angle=90, trim=0 0 0 0]{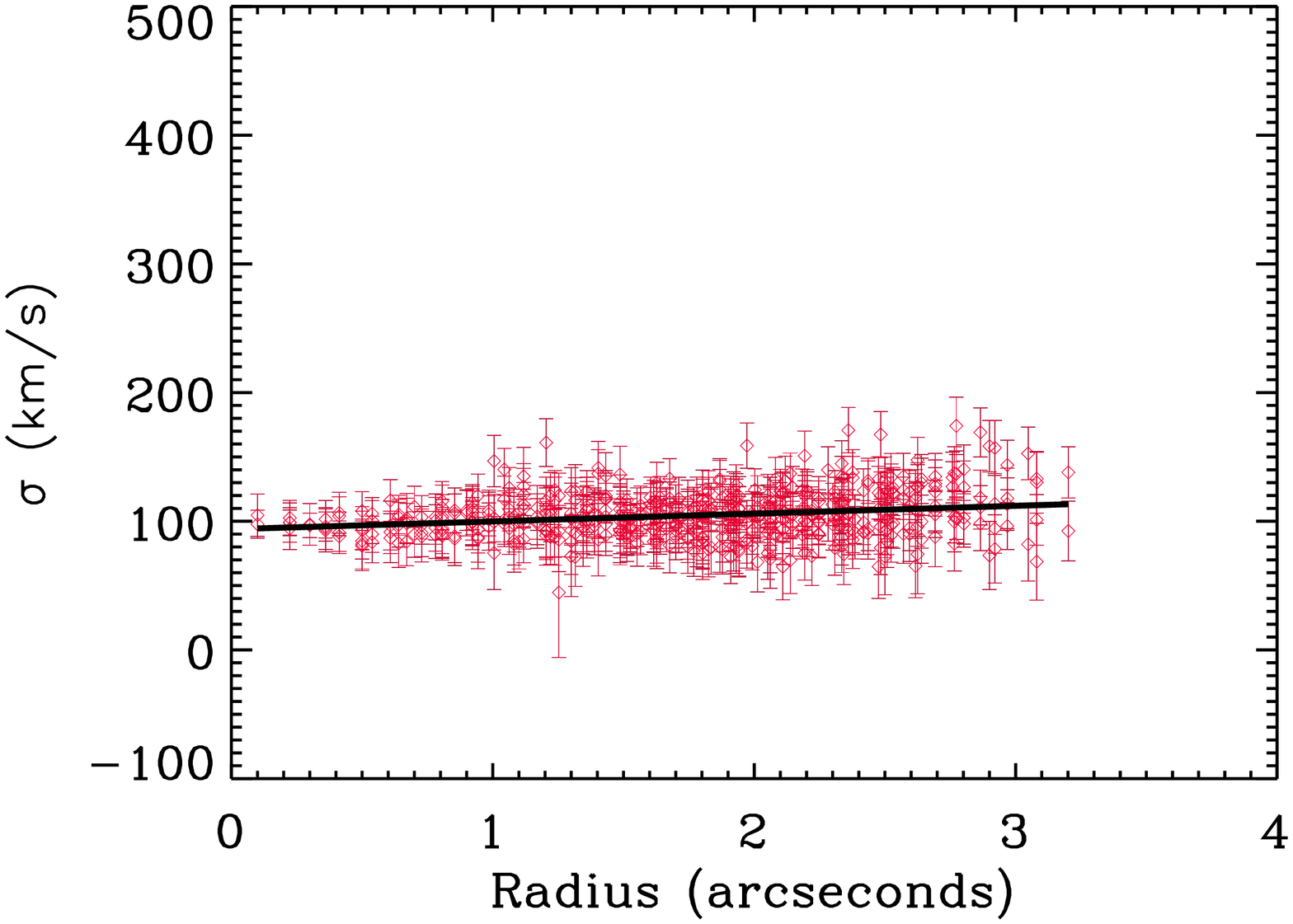}         
    \end{minipage}
\end{center}
    \caption{{\it Left:} two dimensional steaming velocity maps. There is rotation in each case. {\it Right:} measured velocity dispersion within a spaxel
plotted against galacto-centric distance. For the fainter galaxies SJ1613+5103 (1st row) and SJ1718+3007 (third row) the low signal-to-noise ratio results in a large
scatter in the $\sigma$ values and unrealistic values in individual cases. The {\it black line} is a robust fit to the $\sigma$ values as a function of radius and
these fitted values are used in the calculation of the $\lambda_{\rm R}$ parameter to eliminate spurious values dominating the derived value.}
    \label{fig:vel}
\end{figure*}

The stellar population fits described in Section \ref{sec:fit} produce a measurement of the streaming velocity and velocity
dispersion for every spaxel. The two dimensional velocity maps are shown in the {\it left panels} of Figure \ref{fig:vel}. There is rotation
in each case although the maps for SJ1613+5103 (first row) and SJ1718+3007 (third row) are somewhat chaotic. SJ1613+5103 is one member of an
interacting pair and disturbance of the kinematic field is not unexpected. In the case of SJ1718+3007 the kinematic centre appears offset from
the photometric centre (compare with column 2 in Figure \ref{fig:images}). While this may represent evidence of a past merger event which could also
explain the post-starburst spectroscopic signature; we note that these two galaxies are the faintest targets  (see Table \ref{tab:targets}) and have significantly 
lower signal-to-noise ratio than the two brighter galaxies which have smooth rotation fields.

The velocity dispersion measurements are shown
in the {\it right panels} of Figure \ref{fig:vel} plotted as a function of distance from the galaxy centre. These values have large uncertainties when
measured from low signal-to-noise spectra as demonstrated by the scatter and large errors for the velocity dispersion measurements for SJ1613+5103 and SJ1718+3007.
Despite the large scatter in the velocity dispersion measured from an individual spaxel in low signal-to-noise cases, the velocity dispersion as a function
of radius can be well represented by a linear fit. A robust linear fit to the velocity dispersion versus radius is over-plotted as a {\it black line} in Figure \ref{fig:vel}.

The $\lambda_{R}$ parameter \citep{emsellem07,emsellem11} is commonly used to quantify the kinematic state of early type galaxies when two dimensional integral field
spectroscopy is available. 
The $\lambda_{R}$ parameter involves luminosity-weighted averages over the 2D kinematic field provided by IFU data and acts as a proxy to the observed projected stellar angular momentum
per unit mass. $\lambda_{R}$ is defined as:
$$\lambda_{R}= {{\sum_{i=1}^{N} F_{i} R_{i} |V_{i}|} \over {\sum_{i=1}^{N}}F_{i}R_{i}\sqrt{V_{i}^{2}+\sigma_{i}^{2}}}$$
where $F_{i}$, $R_{i}$, $V_{i}$ and $\sigma_{i}$ are the flux, radial distance, streaming velocity and velocity dispersion of the ith spaxel, respectively. The sum runs over all spaxels.

\begin{figure}
      \includegraphics[width=8.8cm, angle=0, trim=80 0 0 0]{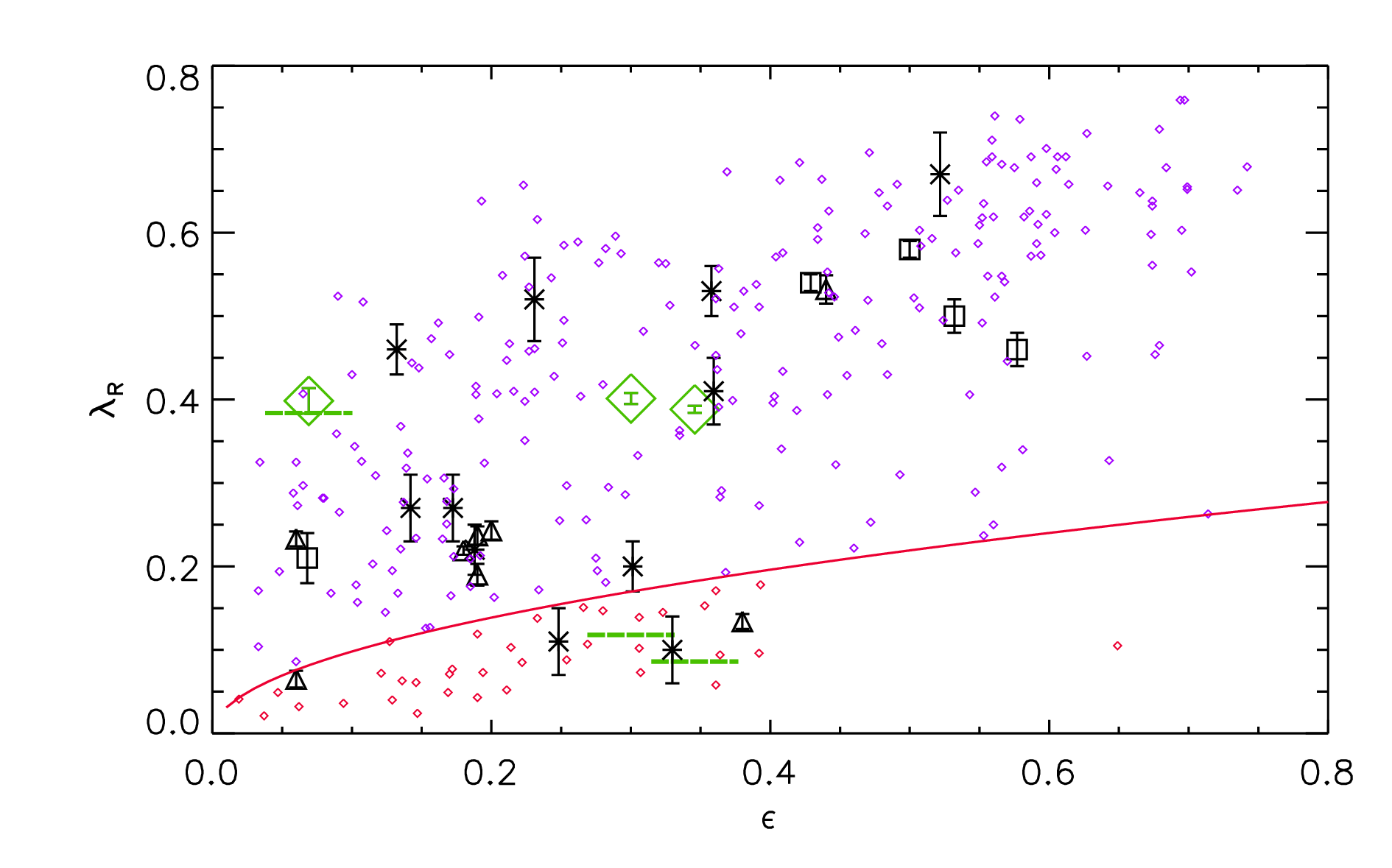}
\caption{\label{fig:lamb}$\lambda_{\rm R}$ versus ellipticity for the three early type galaxies in our sample ({\it green diamonds}). SJ1613+5103
is not plotted since it is a later-type interacting system. The minimum measurable values, as described in Section 3.2, are shown as {\it thick dashed green lines}. The $\lambda_{\rm R}$ values 
previously published are also plotted: \citet{pracy09} {\it black triangles}; \citet{swinbank12} {\it black stars}; \citet{pracy12} {\it black squares}. the ATLAS-3D data \citep{emsellem11} are
overlaid for fast ({\it small blue diamonds}) and slow ({\it small red diamonds}) rotators.}
\end{figure}

\citet{emsellem07} demonstrated  that using this parameter the elliptical galaxy population 
can be separated into two distinct kinematic classes: the fast and slow rotators. \citet{emsellem11}, utilizing the 
complete ATLAS-3D \citep{cappellari11} sample, found $86\pm 2$\,per cent of early type galaxies are 
classified as fast rotators. Likewise, the E+A galaxy population is dominated by fast rotators \citep{pracy09,swinbank12,pracy12}. The dearth of slow rotators in the E+A population
has been used to argue against the need for major galaxy mergers in their production \citep{pracy09,pracy12}, since the probability of a rotating remnant increases as the 
mass ratio of the progenitors involved in the merger increases \citep{bournaud08}. The $\lambda_{\rm R}$ values for our sample are shown in Figure \ref{fig:lamb} as {\it green diamonds}, 
noting we have excluded SJ1613+5103, since it is a later-type interacting system whereas the $\lambda_{\rm R}$ parameter is used as a diagnostic for early type galaxies. 
We use the velocity dispersion values estimated from the linear fit as to exclude spurious values of the velocity dispersion in some low signal-to-noise spaxels.
All three galaxies reside in the fast rotator region of the $\lambda_{\rm R}$--ellipticity plane. 
The errors on $\lambda_{\rm R}$ are estimated by simulating a large number of velocity and velocity dispersion maps by adding a random velocity 
drawn from a normal distribution with a width determined by the error for the velocity or velocity dispersion of that spaxel and recalculating $\lambda_{\rm R}$.

The $\lambda_{\rm R}$ parameter depends on the square of the velocity of each spaxel and as a result is always a positive  quantity. Even in the absence of rotation, noise in the velocity measurements
will result in a non-zero positive value of $\lambda_{\rm R}$ and therefore there is  a minimum value below which the true value of $\lambda_{\rm R}$ cannot be measured. This minimum measurable
value can be estimated by performing the simulations described above after having set the velocity field to zero everywhere. The estimated minimum value of $\lambda_{\rm R}$ from the simulations
can be compared to the $\lambda_{\rm R}$ values measured from the real data. In the case of  SJ2114+0032 and SJ0044-0853, the measured values of  $\lambda_{\rm R}$ are $\sim$3.5 and $\sim$4.5 
times larger than the minimum values and therefore the classification of these galaxies as fast rotators is robust. In the case of SJ1718+3007 (object with the low signal-to-noise velocity 
field in the third row of Figure \ref{fig:vel}) the minimum value and the measured value are essentially the same (within 1$\sigma$), implying 
that the quality of the data are insufficient to classify this galaxy's kinematic state. The same is true for the late type interacting system SJ1613+5103.

In Figure \ref{fig:lamb} we also show previous measurements of $\lambda_{\rm R}$ from the literature. The $\lambda_{\rm R}$ values of the \citet{pracy09} sample are plotted
as {\it black triangles}, the \citet{swinbank12} sample as {\it black stars} and the \citet{pracy12} sample as {\it black squares}. The ATLAS-3D fast rotators ({\it small blue diamonds})
and slow rotators ({\it small red diamonds}) are also shown for comparison \citep{emsellem11}. The {\it thick red curve} is the ellipticity dependent separation of fast rotators (above the line)
and slow rotators (below the line) from \citet{emsellem11}. Of the twenty six E+A galaxies with measured values of  $\lambda_{\rm R}$ (we have excluded SJ1718+3007), four are 
classified as slow rotators using the definition of  \citet{emsellem11}. One of the \citet{pracy12} fast rotator classifications is marginal 
and based on the error bars (they do not quote minimum measurable values) one of the \citet{swinbank12} classifications is also marginal. 
The \citet{pracy09} classifications are robust based on similar simulations to those presented here. 
This equates to a fast rotator fraction
of $\sim$83--85\,per cent. This is similar to the fraction in the overall early type galaxy population as measured from the ATLAS-3D
sample by \citet{emsellem11} of $86\pm 2$\,per cent. This comparison is only illustrative since the $\lambda_{\rm R}$ fraction is sensitive to the 
stellar mass (or luminosity) distribution of sample -- in the sense that there is a higher fraction of slow rotators amongst more massive ellipticals. The distribution 
of absolute K--band magnitudes of the ATLAS-3D sample and the E+A samples are
different. The distribution for the E+A galaxies peaks brighter but extends fainter. In fact, three of the five early type E+A galaxies in the sample of
\citet{pracy12} have absolute K--band magnitudes fainter than the ATLAS-3D selection limit of $M_{\rm k}<-21.5$.

\section{Summary and conclusions}
In this paper we have presented IFU spectroscopy of four E+A galaxies that have a luminosity close to $L_{\rm R}^*$ and overlap
the bottom end of the luminosity distribution of previous studies \citep{pracy09, swinbank12} but which are located at low enough redshift ($z<0.035$) and observed in good enough 
conditions (seeing always $<0.9$\arcsec) to resolve the central kiloparsec -- a condition not met by the previous higher redshift samples. 
Consistent with the properties of the local low luminosity E+A galaxies \citep{pracy12}, we find all four 
have centrally concentrated Balmer line gradients on scales of $\sim 1$\,kpc. This includes the cluster member galaxy SJ0044-0853. 
The gradients correspond to ages of $\lesssim 1$\,Gyr in the centre to a few Gyr at radii of $\sim 1$\,kpc. 
These gradients are consistent with production via galaxy mergers and the Balmer line profiles are similar to those obtained from merger models \citep{pracy05,bekki05}
seen at times of $\sim 0.75$\,Gyr from the peak of the starburst (see Figure \ref{fig:images}).

The ubiquity of the detection of a centrally concentrated young stellar population, and the presence of stellar population gradients 
when the E+A galaxy cores are well resolved, raises the possibility that spatial resolution issues may be important in interpreting the higher redshift results, which
have generally indicated that the post--starburst signature is a more wide spread or global phenomenon. 
We have illustrated this by artificially smoothing our data cubes to simulate them being observed at $z=0.1$ and re-measured the radial H$\delta$ profiles. When this is done the 
observed radial gradients are flattened out to the extent that they  are consistent with no radial gradient at all. However we can not truly simulate the higher redshift observations
since the physical area of the galaxy covered by our observations is smaller than what would be observed (with the same instrument) at higher redshift.
These results are consistent with the models of \citet{pracy10} who demonstrated that the combination of the expected Balmer line gradients in E+A galaxies and the steep surface
brightness profiles of early type galaxies can potentially lead to steep Balmer line gradients being observed as uniformly
high Balmer absorption across the entire galaxy, if the physical scale resolution is insufficient.

A summary of the findings from this study are as follows:
\begin{list}{$\bullet$}{\itemsep=-0.1cm}
\item All four E+A  galaxies have  negative Balmer line gradients in the central $\sim 1$\,kpc. This
is consistent with expectations of formation as the result of galaxy interactions or mergers.\\

\item The Balmer line gradients correspond to ages of $\lesssim 1$\,Gyr in the centre and 
a few Gyrs at galacto-centric radii of $\sim$1\,kpc. \\

\item The three E+A galaxies that are of early morphological type all have values of $\lambda_{\rm R}$
placing them in the fast rotator category, although in one case this classification is not secure as a result
of the low signal-to-noise in the kinematic maps.\\

\item The total number of E+A galaxies with $\lambda_{\rm R}$ measurements is twenty six and four of these
have a slow rotator classification using the ellipticity dependent definition of \citet{emsellem11}. This is close to
the fraction in the early type galaxy population as a whole.\\

\end{list}

\section{Acknowledgments}
Based on observations obtained at the Gemini Observatory, which is operated 
by the Association of Universities for Research in Astronomy, Inc., under a 
cooperative agreement with the NSF on behalf of the Gemini partnership: the 
National Science Foundation (United States), the Science and Technology 
Facilities Council (United Kingdom), the National Research Council 
(Canada), CONICYT (Chile), the Australian Research Council 
(Australia), Minist\'{e}rio da Ci\^{e}ncia, Tecnologia e 
Inova\c{c}\~{a}o (Brazil) and Ministerio de Ciencia, Tecnolog\'{i}a e Innovaci\'{o}n 
Productiva (Argentina). MBP, SC, ES and WJC acknowledge the financial support of the Australian Research 
Council throughout the course of this work. We would like to thank the referee for
insightful comments which greatly improved this paper.


\bsp

\bibliographystyle{mn2e}
\bibliography{references}

\label{lastpage}

\end{document}